\documentclass[fleqn,usenatbib]{mnras}

\usepackage{newtxtext,newtxmath}

\usepackage[T1]{fontenc}
\usepackage{ae,aecompl}

\usepackage{graphicx}	
\usepackage{amsmath}	
\usepackage[dvipsnames]{xcolor}
\usepackage{multirow}
\usepackage{delarray}
\usepackage{color}
\usepackage{here}
\usepackage{float}
\usepackage{tensor}
\usepackage{xspace}
\usepackage{orcidlink}
\usepackage[multiple]{footmisc}

\newcommand{\mach}{\mathcal{M}}

\newcommand{\be}{\begin{equation}} \newcommand{\ee}{\end{equation}}

\newcommand{\SFE}{\mathrm{SFE}}
\newcommand{\solarmass}{\mathrm{M}_{\rm \sun}}
\newcommand{\msun}{\solarmass}
\newcommand{\Msink}{M_\mathrm{sink}}
\newcommand{\Nsink}{N_\mathrm{sink}}
\newcommand{\Mmean}{M_{\rm mean}}
\newcommand{\Mcompleteness}{M_{\rm complete}}
\newcommand{\Mmax}{M_{\rm max}}
\newcommand{\slope}{\gamma_{1,10}}
\newcommand{\Mmedian}{M_{\rm med}}
\newcommand{\tff}{t_\mathrm{ff}}
\newcommand{\tsf}{\tilde{t}}
\newcommand{\epsff}{\epsilon_\mathrm{ff}}
\newcommand{\Mmassmedian}{M_{\rm 50}}
\newcommand{\Msonic}{M_{\rm sonic}}
\newcommand{\MBE}{M_{\rm BE}^{\rm turb}}

\newcommand{\MJeans}{M_{\rm Jeans}}
\newcommand{\alphath}{\alpha_{\mathrm{th}}}
\newcommand{\alphaturb}{\alpha_{\mathrm{turb}}}
\newcommand{\alphaB}{\alpha_{\mathrm{B}}}
\newcommand{\cs}{c_{\rm s}}

\newcommand{\solarluminosity}{\mathrm{L}_{\rm \sun}}
 
\newcommand{\pc}{\mathrm{pc}}
\newcommand{\AU}{\mathrm{AU}}
\newcommand{\kelvin}{\mathrm{K}}

\newcommand{\dderiv}{\mathrm{d}}
\newcommand{\vvector}{\mathbf{v}}

\newcommand{\appropto}{\mathrel{\vcenter{
  \offinterlineskip\halign{\hfil$##$\cr
    \propto\cr\noalign{\kern2pt}\sim\cr\noalign{\kern-2pt}}}}}

\newcommand{\myquote}[1]{``#1''}

\usepackage[normalem]{ulem}

\defcitealias{Guszejnov_isoT_MHD}{Paper I}
\defcitealias{grudic_starforge_methods}{Methods Paper}
\defcitealias{guszejnov_starforge_jets}{Paper II}
\defcitealias{kroupa_imf}{K02}

\title[STARFORGE: What sets the IMF?]{Effects of the environment and feedback physics on the initial mass function of stars in the STARFORGE simulations}

\author[]{
D\'avid Guszejnov\orcidlink{0000-0001-5541-3150}$^{1}$\thanks{guszejnov@utexas.edu},
Michael Y. Grudi\'{c}\orcidlink{0000-0002-1655-5604}$^{2}$\thanks{NASA Hubble Fellow},
Stella S. R. Offner\orcidlink{0000-0003-1252-9916}$^{1}$,
\newauthor
Claude-Andr{\'e} Faucher-Gigu{\`e}re\orcidlink{0000-0002-4900-6628}$^{3}$
Philip F. Hopkins\orcidlink{0000-0003-3729-1684}$^{4}$,
Anna L. Rosen\orcidlink{0000-0003-4423-0660}$^{5}$
\\
$^{1}$Department of Astronomy, University of Texas at Austin, TX 78712, USA \\
$^{2}${Carnegie Observatories, 813 Santa Barbara St, Pasadena, CA 91101, USA}\\
$^{3}${CIERA and Department of Physics and Astronomy, Northwestern University, 1800 Sherman Ave, Evanston, IL 60201, USA}\\
$^{4}$TAPIR, Mailcode 350-17, California Institute of Technology, Pasadena, CA 91125, USA \\
$^{5}$Center for Astrophysics $|$ Harvard \& Smithsonian, 60 Garden St, Cambridge, MA 02138, USA \\
}

\date{\today \vspace{-0.6cm}}

\pubyear{2022}

\begin{document}
\label{firstpage}
\pagerange{\pageref{firstpage}--\pageref{lastpage}}
\maketitle

\begin{abstract}
One of the key mysteries of star formation is the origin of the stellar initial mass function (IMF). The IMF is observed to be nearly universal in the Milky Way and its satellites, and significant variations are only inferred in extreme environments, such as the cores of massive elliptical galaxies and the Central Molecular Zone. In this work we present simulations from the STARFORGE project that are the first cloud-scale radiation-magnetohydrodynamic simulations that follow individual stars and include all relevant physical processes. The simulations include detailed gas thermodynamics, as well as stellar feedback in the form of protostellar jets, stellar radiation, winds and supernovae. In this work we focus on how stellar radiation, winds and supernovae impact star-forming clouds. Radiative feedback plays a major role in quenching star formation and disrupting the cloud, however the IMF peak is predominantly set by protostellar jet physics. We find the effect of stellar winds is minor, and supernovae \myquote{occur too late} to affect the IMF or quench star formation. We also investigate the effects of initial conditions on the IMF. We find the IMF is insensitive to the initial turbulence, cloud mass and cloud surface density, even though these parameters significantly shape the star formation history of the cloud, including the final star formation efficiency. Meanwhile, the characteristic stellar mass depends weakly on metallicity and the interstellar radiation field, which essentially set the average gas temperature. Finally, while turbulent driving and the level of magnetization strongly influences the star formation history, they only influence the high-mass slope of the IMF.
\end{abstract}



 \section{Introduction}\label{sec:intro}

Although star formation is one of the most fundamental processes of astrophysics,  there is no widely accepted theory of star formation, despite decades of intensive work from both observers and theorists \citep{mckee_star_formation, sf_big_problems}. The primary reason for this is the large set of interconnected, complex physical processes, including gravity, turbulence, magnetic fields, chemistry and radiation \citep{Girichidis_2020_sf_processes_review}. Furthermore, these processes interact in a non-linear way that also create interactions between vastly different scales (e.g. feedback from massive stars affecting their progenitor cloud). Thus, in order to understand star formation it is vital to investigate the role each physical process plays and how it modifies the outcome. A key question is how these processes affect the at-formation stellar mass spectrum, i.e., the initial mass function (IMF). 
 
Since the large set of complex physical processes involved prevents a direct treatment, star formation models (both analytical and numerical) can only include a limited subset of physical processes. Analytic models and early simulations modeled the dense, star forming clouds of the Milky Way as isothermal, turbulent objects collapsing under self-gravity \citep[e.g.,][]{padoan_nordlund_2002_imf,hc08,core_imf}. Recent numerical works have shown that this set of physics is inadequate to produce a converged mass spectrum for collapsed fragments (i.e. stars, see \citealt{Martel_numerical_sim_convergence,Kratter10a,guszejnov_feedback_necessity,Federrath_2017_IMF_converge_proceedings, guszejnov_isothermal_collapse, Lee_Hennebelle_2018_EOS}), so additional physics must play a role.

Star-forming clouds are observed to have significant support from magnetic fields \citep{crutcher_2009_mc_magnetic_fields}. Both theoretical and numerical works have found that the addition of magnetic fields impose a resolution independent scale on the stellar mass spectrum \citep[see e.g.][]{padoan_2007, padoan_nordlund_2011_imf, Haugbolle_Padoan_isot_IMF}. Several of these studies claimed to reproduce the observed IMF. However, a larger parameter study of high-resolution simulations (\citealt{Guszejnov_isoT_MHD}, henceforth referred to as \citetalias{Guszejnov_isoT_MHD}) showed that the mass scale imposed by magnetic fields is both too large and too sensitive to initial conditions, in a way that would violate the observed near-universality of the IMF \citep{imf_review}. 

While clouds are close to isothermal for a wide range of densities, there can be significant deviations even at low densities around $\sim 10^2\,\mathrm{cm}^{-3}$ (see \citealt{Glover_Clark_n_T}). In high-density regions ($\sim 10^5\,\mathrm{cm}^{-3}$) the isothermal assumption breaks down as the cloud becomes opaque to its own cooling radiation, leading to increased temperatures and suppressed fragmentation \citep{lowlyndenbell1976, rees1976,Lee_Hennebelle_2018_EOS,Colman_Teyssier_2019_tidal_screening}. In our recent work (\citealt{guszejnov_starforge_jets}, henceforth referred to as \citetalias{guszejnov_starforge_jets}), we showed that non-isothermal effects by themselves are insufficient to lower the characteristic stellar mass to the observed value, and these effects are most significant in providing a minimum mass scale for star formation (see \citealt{lowlyndenbell1976}).

Recent works show that the energy and momentum injected by newly formed stars to their environment (i.e. stellar feedback) can dramatically affect the star formation process \citep{Offner_2009_radiative_sim, krumholz_stellar_mass_origin, bate12a, Myers_2013_ORION_radiation_IMF,guszejnov_gmc_imf,guszejnov_feedback_necessity,Rosen_Offner_2020_high_mass_stars}. The first of these processes to act during star formation is the launching of high-velocity bipolar outflows by accreting protostars \citep[see reviews of][]{Frank_2014_jets_preview, Bally_2016_outflows_review} that are likely driven by collimated bipolar jets launched along the rotational axis of the accreting protostar \citep{Rosen_krumholz_2020_outflows_massive_stars}, which are produced by the magnetic interaction between the protostar and its accretion disk \citep{Shu_1988_X_winds,Pelletier_1992_D_wind} and radiation pressure in the case of massive protostars \citep{Kuiper_2010_massive_star_radiation_outflow,Vaidya_2011_massive_jet_sim, Rosen_2016_massive_star_RT}. Previous work has shown that these jets both reduce accretion rates and drive turbulence on small scales \citep{Nakamura_2007_outflow_turbulence_driving,Matzner_2007_outflow_tub_obs,Wang_2010_outflow_regulated_SF,Cunningham_2011_outflow_sim,Offner_Arce_2014,Federrath_2014_jets,Offner_Chaban_2017_jets_sfe, murray_2018_jets,Rohde_2021_outflows_IMF_multiplicities, Appel_2022_outflow_effect_on_density_PDF}. Recently in \citetalias{guszejnov_starforge_jets} we showed that jets can dramatically lower the characteristic stellar mass as they disrupt the accretion flow around stars, allowing the nearby gas to fragment \citep[similar to the results of][]{li_2018_sf_mhd_jets,Cunningham_2018_feedback, Mathew_Federrath_2021_IMF_multiplicity_2021}. Overall jets reduced stellar masses by an order of magnitude compared to MHD simulations without stellar feedback, thereby bringing the simulated stellar mass spectrum in agreement with the observed IMF. However, \citetalias{guszejnov_starforge_jets} also found that jets by themselves are insufficient to regulate star formation and that massive stars undergo runaway accretion without additional feedback processes.

Radiation from accreting protostars is thought to be a crucial ingredient of the star formation process for both low and high-mass stars \citep{Offner_2009_radiative_sim,krumholz11a, krumholz_stellar_mass_origin, bate12a, Myers_2013_ORION_radiation_IMF,guszejnov_gmc_imf,guszejnov_feedback_necessity, Rosen_2016_massive_star_RT, Cunningham_2018_feedback, Rosen_Offner_2020_high_mass_stars} as they can heat their surroundings, preventing fragmentation. Once stars reach the main sequence they start emitting ionizing radiation as well as isotropic line-driven stellar winds, both of which can dramatically affect their surroundings, potentially halting stellar accretion \citep{krumholz_2012_orion_sims,Rosen_2016_massive_star_RT, li_2018_sf_mhd_jets,Cunningham_2018_feedback,Rosen_Offner_2020_high_mass_stars, rosen_MSF_winds2022}. Massive stars in particular provide feedback powerful enough to affect the IMF in their entire natal cloud \citep{Gavagnin_2017_SF_feedback}, as well as completely quench star formation (see \citealt{krumholz_2019_cluster_review} for review and Fig. 1  of \citealt{grudic_mond} for a literature compilation of theoretical predictions), helping to limit the star formation efficiency of clouds to a few percent \citep{grudic_2016, kim_2018_gmc_raytrace,grudic_2018_mwg_gmc, Li_Vogelsberger_2019_GMC_disrupt, grudic_2020_cluster_formation}.

The lifetime of massive stars is several Myr, comparable to the lifetime of star-forming clouds. The resulting supernovae (SNe) dominate the momentum input by stellar feedback in the ISM and are critical to regulate star formation on galactic scales \citep{Somerville_Dave_2015_galaxy_formation_review, naab_ostriker_galform_review, vogelsberger_galform_review} and could be a vital ingredient to the cloud-scale star formation process as well. Their effect, however, is reduced by the fact that they act \myquote{late} in the star formation process, so it is unclear how much they either affect the IMF or regulate star formation. Simulations of star-cluster formation have found they have negligible impact upon star formation efficiency and bound cluster masses compared to early feedback (i.e. radiation), even in massive GMCs that survive long enough to host a SN before disruption \citep{grudic_2020_cluster_formation}. Nevertheless, SNe must at least play an indirect role in cloud-scale star formation as they are thought to be one of the main drivers of galactic turbulence and thus set the properties of GMCs \citep[e.g.,][]{ostriker_shetty_sf_turbulence_sne,Hopkins_2011_SFR_self_regulate,hopkins_2012_galaxy_structure,Faucher_Giguere_2013_SF_feedback_galactic_disks, Padoan_2017_SN_driving_SFE,Seifried_2018_GMC_SN_driving, guszejnov_GMC_cosmic_evol,Gurvich_2020_ISM_pressure_balance__gal_disks}.

Simulations that take into account the above processes are necessary to understand the effects of each physical process, but so far such studies have generally been limited to simple physics or to a very narrow range of cloud initial conditions. In this paper we introduce a suite of results from the STAR FORmation in Gaseous Environments (STARFORGE) project\footnote{\url{http://www.starforge.space}} that include all of the above physical processes. These radiation-magnetohydrodynamic (RMHD) simulations achieve a dynamic range in mass resolution that is an order of magnitude higher than any previous star cluster simulation, allowing us to simulate the detailed evolution of molecular clouds while following the formation of individual stars with stellar masses as low as $\sim$0.1~$M_{\rm \odot}$ (see methods paper of \citealt{grudic_starforge_methods}, henceforth referred to as \citetalias{grudic_starforge_methods}). In this study we perform and analyze a set of simulations with different initial conditions (ICs) and levels of physics to identify the impact of outflows, stellar radiation, winds and supernovae on the IMF and the star formation history of clouds. 
 
First, we provide a brief overview of the STARFORGE codebase and the various parameters and metrics we use in \S\ref{sec:methods} (for a more detailed discussion see the \citetalias{grudic_starforge_methods}). We present our results in \S\ref{sec:results_ladder} with a focus on how the star formation history and the characteristic masses of sink particles (stars) change with the inclusion of additional physics. In \S\ref{sec:results_sensitivity} we explore variations in the initial conditions (e.g., cloud surface density, level of turbulence) and physical parameters (e.g., turbulent driving). The implications of these results as well as the potential role of further, not-yet included physics are discussed in \S\ref{sec:discussion}. We summarize our conclusions in \S\ref{sec:conclusions}. 

Note that for brevity we are only showing figures essential for the main points of this paper. For additional figures we refer the reader to the online supplementary materials of this paper, which can also be found in a GitHub repository\footnote{\url{https://github.com/guszejnovdavid/STARFORGE_IMF_paper_extra_plots}}.


 \section{Numerical Methods}\label{sec:methods}
 
  \subsection{The STARFORGE simulations}\label{sec:starforge}
  
  For this work we utilize simulations from the STARFORGE project, which are run with the {\small GIZMO} simulation code\footnote{\url{http://www.tapir.caltech.edu/~phopkins/Site/GIZMO.html}}. A full description and presentation of the STARFORGE methods including a variety of tests and algorithm details are given in the \citetalias{grudic_starforge_methods}, therefore we only briefly summarize the key points here. Readers familiar with the STARFORGE simulations should skip ahead to \S\ref{sec:results_ladder}.
  
  \subsubsection{Physics}\label{sec:physics}
  
  We simulate star-forming clouds with the {\small GIZMO} code \citep{hopkins2015_gizmo}, using the Lagrangian meshless finite-mass (MFM) method for magnetohydrodynamics \citep{hopkins_gizmo_mhd}, assuming ideal MHD. 
  Sink particles represent individual stars. Once they form they follow the protostellar evolution model from \citet{Offner_2009_radiative_sim}.

  \myquote{Non-isothermal} or \myquote{cooling} STARFORGE runs utilize the radiative cooling and thermochemistry module from \citet{fire3} that contains detailed metallicity-dependent cooling and heating physics from $T=10-10^{10}\,$K, including recombination, thermal bremsstrahlung, metal lines (following \citealt{Wiersma2009_cooling}), molecular lines, fine structure (following \citealt{Glover_Abel_2008}) and dust collisional processes. The cooling module self-consistently solves for the internal energy and ionization state of the gas (see \citet{fire3} and Appendix B of \citealt{hopkins2017_fire2}). 
  STARFORGE simulations use two different treatments of radiation transport, in this work we present \myquote{RHD} simulations that co-evolve the gas, dust, and radiation temperature self-consistently (unlike in \citetalias{guszejnov_starforge_jets}), including the stellar luminosity in various bands accounting for photon transport, absorption and emission using dust opacity. 
  In addition to local sources (i.e. stars) we include an external heating source that represents the interstellar radiation field (ISRF).

  As shown in \citetalias{guszejnov_starforge_jets} protostellar jets represent a crucial feedback mechanism. We model their effects by having sink particles launch a fixed fraction ($f_w=0.3$) of the accreted material along their rotational axis with a fixed fraction ($f_K=0.3$) of the Keplerian velocity at the protostellar radius.
  
  In addition to their radiative feedback massive main-sequence stars inject a significant amount of momentum and energy into their surroundings through stellar winds. We calculate the mass-loss rates and wind velocities based on \citet{smith_2014_winds} and \citet{lamers_1995_wind_vesc} respectively. Winds are implemented either through local mass, momentum and energy injection or direct gas cell spawning, while SNe are spawned at the end of the lifetime of all $>8\msun$ stars.
  
\subsubsection{Cloud parameters}\label{sec:cloud_params}

To describe our initial conditions we introduce several parameters, using the same definitions as in \citetalias{guszejnov_starforge_jets} and \citetalias{Guszejnov_isoT_MHD}. First, we introduce the \emph{3D sonic Mach number}
\begin{equation}
\mach^2\equiv\langle ||\vvector_\mathrm{turb}||^2/\cs^2\rangle,
\label{eq:mach}
\end{equation}
where $\cs$ is the gas sound speed and $\vvector_\mathrm{turb}$ is the turbulent velocity field, while $\langle...\rangle$ denotes mass-weighted averaging. It is also useful to introduce the \emph{turbulent virial parameter} $\alphaturb$, which measures the relative importance of turbulence to gravity, following the convention in the literature (e.g., \citealt{Bertoldi_McKee_1992, federrath_sim_2012}),
\begin{equation}
\alphaturb \equiv \frac{5 ||\vvector_\mathrm{turb}||^2 R_\mathrm{cloud}}{3 G M_\mathrm{0}} = \frac{5 \mach^2 \cs^2 R_\mathrm{cloud}}{3 G M_\mathrm{0}},
\label{eq:alphaturb_sphere}
\end{equation} 
where $R_\mathrm{cloud}$ and $M_\mathrm{0}$ are the cloud (spherical-equivalent) radius and total mass. 
The relative importance of the magnetic field is commonly described by the \emph{normalized magnetic flux} (or mass-to-flux ratio), which for a uniform magnetic field can be expressed as:
\begin{equation}
\mu = c_1\sqrt{\frac{-E_{\rm grav}}{E_{\rm mag}}},
\label{eq:mu}
\end{equation}
where the normalization constant $c_1\approx 0.4$ \citep{Mouschovias_Spitzer_1976_magnetic_collapse}. 


For a detailed definition of these quantities and the others listed in Table \ref{tab:IC_phys} see \S2 in \citetalias{Guszejnov_isoT_MHD}.

\subsubsection{Initial Conditions}\label{sec:initial_conditions}
    
  We generate our initial conditions (ICs) using {\small MakeCloud}\footnote{\url{https://github.com/mikegrudic/MakeCloud}}, identical to \citetalias{guszejnov_starforge_jets}. Unless otherwise specified our runs utilize \emph{\myquote{Sphere} ICs}, meaning that we initialize a spherical cloud (radius $R_\mathrm{cloud}$ and mass $M_\mathrm{0}$) with uniform density, surrounded by diffuse gas with a density contrast of 1000. The cloud is placed at the center of a periodic $10 R_\mathrm{cloud}$ box. The initial velocity field is a Gaussian random field with power spectrum $E_k\propto k^{-2}$ \citep{ostriker_2001_mhd}, scaled to the value prescribed by $\alphaturb$. The initial clouds have a uniform $B_z$ magnetic field whose strength is set by the parameter $\mu$. There is no external driving in these simulations. Although the gas is initialized at $T=10\,\kelvin$, but the gas-dust mixture quickly reaches equilibrium with the ISRF, for which we assume solar neighborhood conditions. Also, the gas is initially fully atomic and reaches an equilibrium molecular fraction by the time star formation begins 
  (note that in \textit{Box} runs the equilibrium is reached during the \myquote{stirring} phase).

  We also run simulations using \emph{\myquote{Box} ICs}, similar to the driven boxes used in e.g., \citet{Federrath_2014_jets,Cunningham_2018_feedback}. These are initialized as a constant density, zero velocity periodic cubic box with the same temperature prescription as the \myquote{Sphere} ICs. This periodic box is then \myquote{stirred} using the driving algorithm from \citealt{federrath_sim_compare_2010,bauerspringel2012}. This involves a spectrum of $E_k\propto k^{-2}$ of driving modes in Fourier space at wavenumbers 1/2 - 1 times the box size, with an appropriate decay time for driving mode correlations ($t_{\mathrm{decay}}\sim t_{\mathrm{cross}}$). This stirring is initially performed without gravity for five global freefall times ($\tff$, see Eq. \ref{eq:tff}), to achieve saturated MHD turbulence. The normalization of the driving spectrum is set so that in equilibrium the gas in the box has a turbulent velocity dispersion that gives the desired Mach number $\mach$ and virial parameter $\alphaturb$. We use purely solenoidal driving, which remains active throughout the simulation after gravity is switched on, unless specified otherwise. We take the box side length $L_\mathrm{box}$ to give a box of equal volume to the associated {\it Sphere} cloud model. An important difference between the \textit{Sphere} and \textit{Box} runs is that in the case of driven boxes the magnetic field is enhanced by a turbulent dynamo \citep{federrath_2014_dynamo} and saturates at a relative magnetic energy level of $\alphaB\sim 0.1$ (see \citetalias{Guszejnov_isoT_MHD}), so for Box runs the \myquote{pre-stirring} magnetic field strength (defined by the normalized flux $\mu$) does not directly specify the actual initial magnetic field strength when gravity is turned on (however the \myquote{pre-stirring} flux in the box will still affect the large-scale geometry of the magnetic field). 
  
  Table \ref{tab:IC_phys} shows the target parameters for the runs we present in this paper. The input parameters are the cloud mass $M_0$, radius $R_0$, turbulent virial parameter $\alphaturb$ and normalized magnetic mass-to-flux ratio $\mu$. Similar to \citetalias{guszejnov_starforge_jets} we set up our clouds to lie along a mass-size relation similar to observed GMCs in the Milky Way (e.g. \citealt{larson_law}, specifically assuming $\Sigma\equiv M_\mathrm{0}/ \uppi R_\mathrm{cloud}^2 = 63 \msun\,\mathrm{pc}^{-2}$). These clouds are marginally bound ($\alphaturb=2$) and start out at either $T=10\,\kelvin$ or in equilibrium with the ISRF (RHD runs only). For the initial magnetization we assume $\alphaB=-2E_\mathrm{mag}/E_\mathrm{grav}=0.02$, which translates to $\mu=0.4$. Note that due to the much higher computational cost of RHD runs, only clouds up to $2\times10^4\,\msun$ are simulated with explicitly evolved radiation. Also, in runs where the star formation is not quenched (i.e. those without stellar radiation) we restrict ourselves to times when the star formation efficiency ($\SFE=M_{\star}/M_{0}$) is below 10\% since most MW GMCs achieve a star formation efficiency ($\SFE=M_{\star}/M_{0}$) of 1\%-10\% over their lifetime (see \citealt{sf_big_problems} for a discussion, and note that some clouds have <1\%, see \citealt{Federrath_density_distrib}). Note that the STARFORGE simulations have an effective mass resolution of $\Delta m=10^{-3}\,\msun$, making the mass function incomplete for $M\lesssim \Mcompleteness=0.1\,\msun$, which are thus omitted from our analysis. See Appendix \ref{app:completeness} for a detailed explanation for our choice of $\Mcompleteness$.

\begin{table*}
    \setlength\tabcolsep{2.0pt} 
	\centering
	\begin{tabular}{ | c | c | c | c | c | c | }
	\hline
	\textbf{Physics label} & Thermodynamics  & MHD & Protostellar Jets & Stellar Radiation & Stellar Winds \& SNe \\
	\hline
	\multicolumn{6}{|c|}{\bf \myquote{Physics ladder} of star formation} \\
	\hline
	\textbf{I\_M} & Isothermal (I) & Ideal (M) & \multicolumn{3}{c|}{Not included}  \\
	\hline
	\textbf{C\_M} & Non-isothermal, RHD (RHD) & Ideal (M) & \multicolumn{3}{c|}{Not included}   \\
	\hline
	\textbf{C\_M\_J} & Non-isothermal, RHD (RHD) & Ideal (M) & Included (J) & \multicolumn{2}{c|}{Not included}  \\ \hline
	\textbf{C\_M\_J\_R} & Non-isothermal, RHD (RHD) & Ideal (M) & Included (J) & Included (R) & Not included   \\ \hline
	\textbf{C\_M\_J\_R\_W} & Non-isothermal, RHD (RHD) & Ideal (M) & Included (J) & Included (R) & Included (W)   \\
	\hline
	\multicolumn{6}{|c|}{\bf Physics variation tests} \\
	\hline
	\textbf{ISRFx10} & \multicolumn{5}{c|}{Includes all like C\_M\_J\_R\_W, but the background ISRF is 10 times the solar circle value}  \\
	\hline
	\textbf{ISRFx100} & \multicolumn{5}{c|}{Includes all like C\_M\_J\_R\_W, but the background ISRF is 100 times the solar circle value}  \\
	\hline
	\textbf{Z01} & \multicolumn{5}{c|}{Includes all like C\_M\_J\_R\_W, but metallicity is 10\% of the solar value }  \\
	\hline
	\textbf{Z001} & \multicolumn{5}{c|}{Includes all like C\_M\_J\_R\_W, but metallicity is 1\% of the solar value }  \\
 	\hline
    \end{tabular}
	\begin{tabular}{|cccccc|ccccccccc|c c|}
	     \multicolumn{17}{c}{}\\ 
		 \multicolumn{1}{c}{}&
		 \multicolumn{6}{c}{\bf Input Parameters} &
		 \multicolumn{8}{c}{\bf Derived Parameters}&
		 \multicolumn{2}{c}{\textbf{Resolution}} \\
		\hline
		\bf Cloud label & $M_0$ [$\msun$] & $R_{\mathrm{cloud}}$ [pc] & $L_{\mathrm{box}}$ [pc] & $\alphaturb$ & $\mu$ & $\sigma$ [km/s]  &  $\alphath$ & $\alpha$ & $\mach_{\rm A} $ & $\beta$ & $\alphaB$ & $\frac{\MJeans}{M_0}$ & $\frac{\Msonic}{M_0}$ & $\frac{M_{\Phi}}{M_0}$ &  $M_0/\Delta m$ &  $\Delta x_\mathrm{J}$ [AU] \\
		\hline
		\multicolumn{17}{|c|}{\bf MW cloud analogues} \\
		\hline
		\bf M2e2 & $2\times 10^2$ & 1 &  & 2 & 4.2 & 
		1.0 & 0.02 & 2.04 & 10 &  7.8 & 0.02 & 
		$6\times 10^{-2}$ & $7 \times 10^{-3}$ & 0.1 &  $2\times10^{5}$ & 36  \\
		\hline
		\bf M2e3 & $2\times 10^3$ & 3 & & 2 & 4.2 & 
		1.9 & 0.02 & 2.04 & 10 &  2.3 & 0.02 & 
		$1\times 10^{-2}$ & $6 \times 10^{-4}$ & 0.1 &  $2\times10^{7}$ & 3.6 \\
		\hline
		\bf M2e4 & $2\times 10^4$ & 10 & 16 & 2 & 4.2
		& 3.2 & 0.008 & 2.03 & 10 &  0.78 & 0.02 & 
		$3\times 10^{-3}$ & $7 \times 10^{-5}$ & 0.1 &  $2\times10^{7}$ & 36 \\
		\hline
		\multicolumn{17}{|c|}{\bf Parameter variation tests} \\
		\hline
		\bf M2e4\_R3 & $2\times 10^4$ & 3 &  & 2 & 4.2
		& 5.8 & 0.008 & 2.02 & 10 &  0.23 & 0.02 & 
		$5\times 10^{-4}$ & $7 \times 10^{-6}$ & 0.1 &  $2\times10^{7}$ & 36 \\
		\hline
		\bf M2e4\_R30 & $2\times 10^4$ & 30 &  & 2 & 4.2
		& 1.9 & 0.02 & 2.04 & 10 &  2.3 & 0.02 & 
		$1\times 10^{-2}$ & $6 \times 10^{-4}$ & 0.1 &  $2\times10^{7}$ & 36 \\
		\hline
		\bf M2e4\_a1 & $2\times 10^4$ & 10 &  & 1 & 4.2
		& 2.3 & 0.008 & 1.03 & 10 &  0.78 & 0.02 & 
		$3\times 10^{-3}$ & $4 \times 10^{-5}$ & 0.1 &  $2\times10^{7}$ & 36 \\
		\hline
		\bf M2e4\_a4 & $2\times 10^4$ & 10 &  & 4 & 4.2
		& 4.5 & 0.008 & 4.03 & 10 &  0.78 & 0.02 & 
		$3\times 10^{-3}$ & $1 \times 10^{-4}$ & 0.1 &  $2\times10^{7}$ & 36 \\
		\hline
		\bf M2e4\_mu1.3 & $2\times 10^4$ & 10 &  & 2 & 1.3
		& 3.2 & 0.008 & 2.21 & 3.1 &  0.078 & 0.2 & 
		$3\times 10^{-3}$ & $7 \times 10^{-5}$ & 0.4 &  $2\times10^{7}$ & 36 \\
		\hline
		\bf M2e4\_mu0.4 & $2\times 10^4$ & 10 &  & 2 & 0.42
		& 3.2 & 0.008 & 4.01 & 3.1 &  0.0078 & 2 & 
		$3\times 10^{-3}$ & $7 \times 10^{-5}$ & 4 &  $2\times10^{7}$ & 36 \\
		\hline

	\end{tabular}
        \vspace{-0.1cm}
 \caption{Simulations used in this paper described with STARFORGE label conventions. \textit{Top}: Physics modules included, see \S\ref{sec:physics} and \citetalias{grudic_starforge_methods} for details on the individual physics modules. \textit{Bottom}: Initial conditions of clouds used in our runs, with $M_0$, $R_{\mathrm{cloud}}$, $\alphaturb$ and $\mu$ being the initial cloud mass, size, virial parameter, mass to magnetic flux ratio and temperature respectively (note that the initial gas-dust temperature is set by ISRF). We also report the initial 3D turbulent velocity dispersion $\sigma$, thermal virial parameter $\alphath$ assuming $T=10\,\kelvin$, total virial parameter $\alpha$, Alfv\'{e}n Mach number $\mach_{\rm A}$, plasma $\beta$, magnetic virial parameter $\alphaB$, as well as the relative Jeans, sonic and magnetic mass scales (see \S2 in \citetalias{Guszejnov_isoT_MHD} for definitions). Note that the parameters in this table apply to both \textit{Box} and \textit{Sphere} runs as they are set up to have identical initial global parameters, with  $L_{\mathrm{box}}$ being the box size for \textit{Box} runs and $R_{\mathrm{cloud}}$ being the cloud radius for \textit{Sphere} runs. Note that \textit{Box} runs have slightly different initial parameters (e.g., Mach number, virial parameter) due to the non-exact scaling of the driving, so the values shown here are the target values.}
 \label{tab:IC_phys}\vspace{-0.5cm}
\end{table*}

\begin{figure*}
\begin {center}
\includegraphics[width=0.99\linewidth]{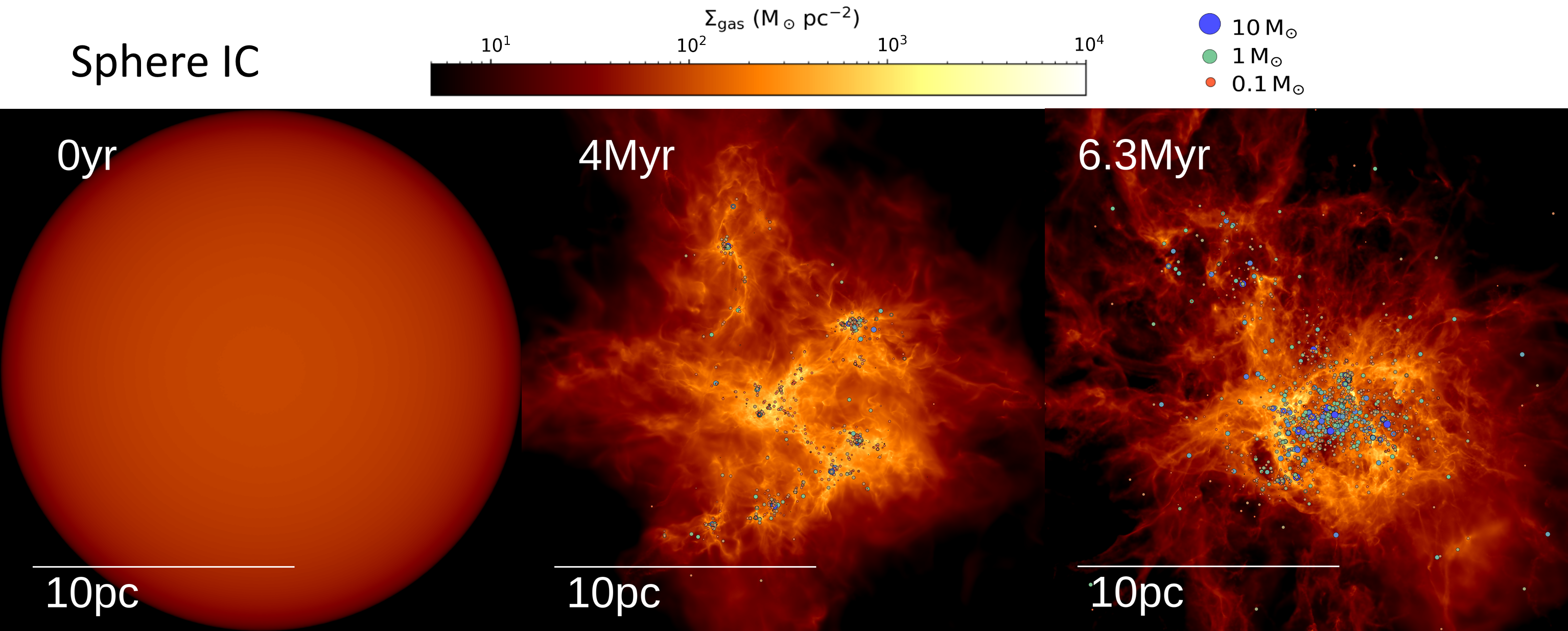}
\\
\includegraphics[width=0.99\linewidth]{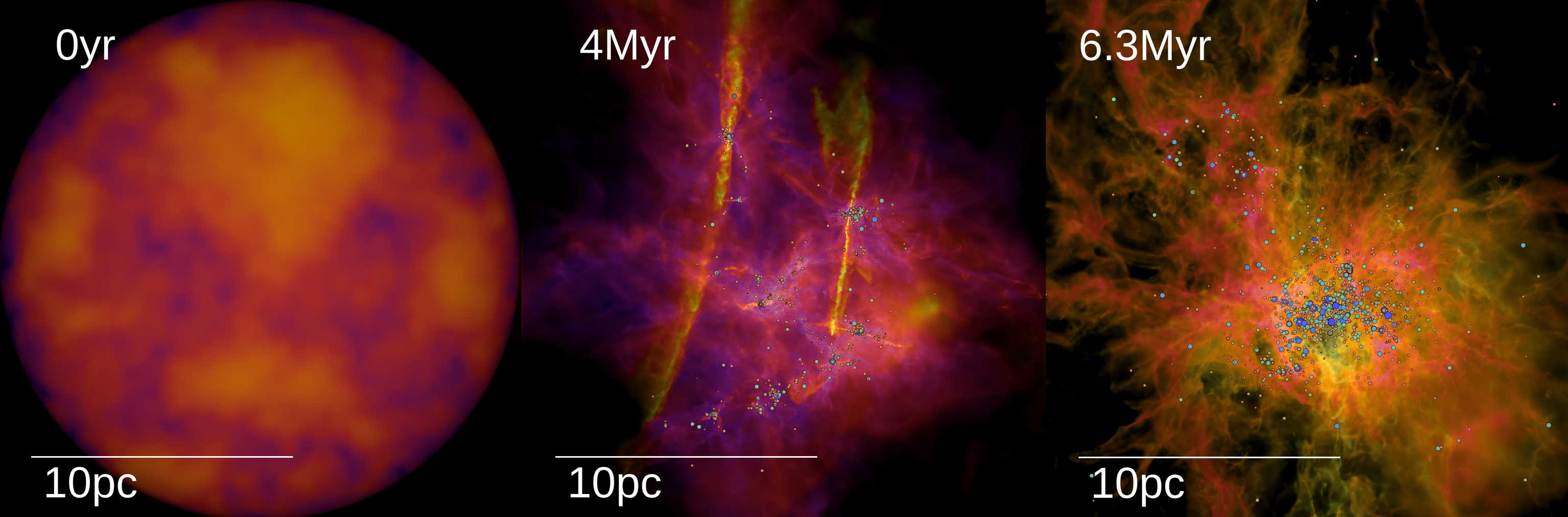}
\vspace{-0.3cm}
\caption{\textit{(Row 1)} Surface density maps for \textbf{M2e4\_C\_M\_J\_RT} with $M_0/\Delta m=2\times 10^7$ initial gas cells (see Table \ref{tab:IC_phys}) at different times for the Sphere IC. The color scale is logarithmic and the circles represent sink particles (stars) that form in high-density regions where fragmentation can no longer be resolved, their size increasing with mass as well as their color changing from red ($M\sim0.1\,\msun$) to blue ($M\sim10\,\msun$). This simulation resolves a dynamic range from $\sim\!\mathrm{20\,pc}$ down $\sim\!\mathrm{30\,AU}$ and is run until stellar feedback quenches star formation and disrupts the cloud (see right column). \textit{(Row 2)} Same as above, but now shown with a color map that encodes the 1D line-of-sight velocity dispersion (increasing from purple ($0.1\rm km\,s^{-1}$) to orange ($10\rm km\,s^{-1}$) and encodes surface density information in lightness (lighter is denser). These kinematic maps can highlight feedback processes that would be invisible in surface density maps (i.e. protostellar jets).
}
\label{fig:M2e4_series}
\vspace{-0.5cm}
\end {center}
\end{figure*} 

\begin{figure*}
\begin {center}
\includegraphics[width=0.99\linewidth]{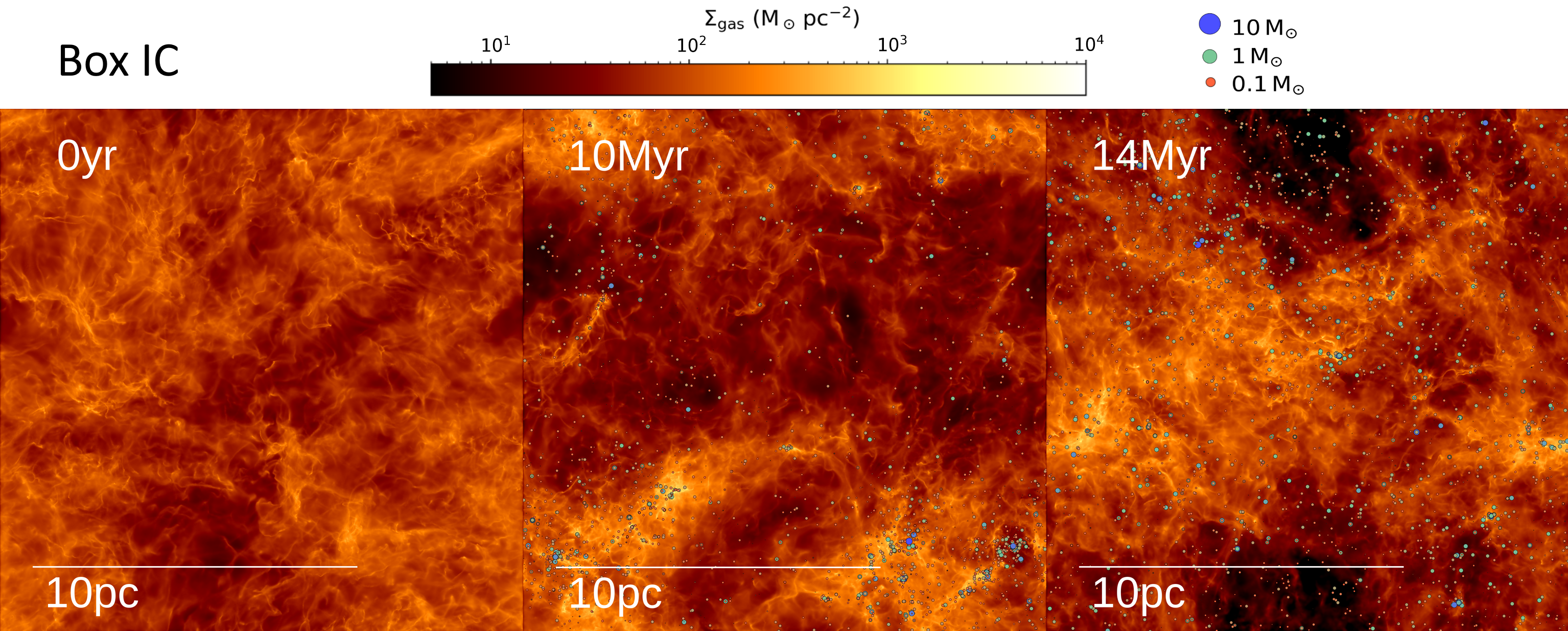}
\\
\includegraphics[width=0.99\linewidth]{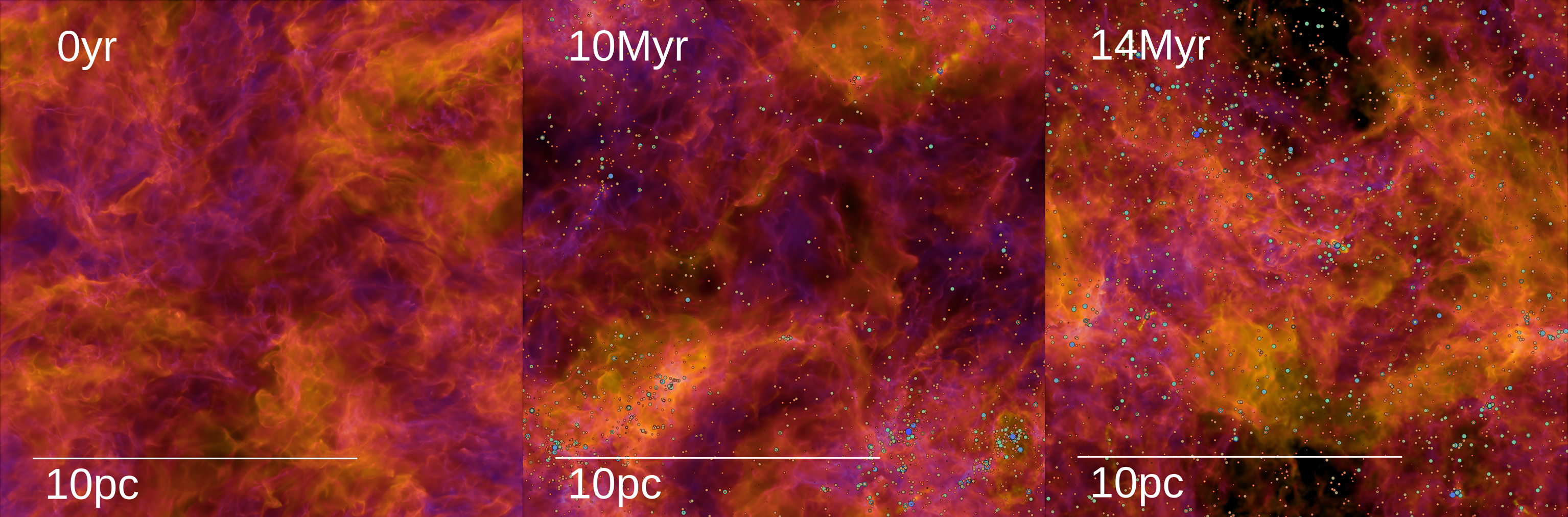}
\vspace{-0.3cm}
\caption{Same as Figure \ref{fig:M2e4_series} but for Box IC.}
\label{fig:M2e4_series_Box}
\vspace{-0.5cm}
\end {center}
\end{figure*} 

\subsubsection{Star formation metrics}\label{sec:metrics}

To describe the different aspects of the star formation process in our simulations we use a series of variables, including the star formation efficiency $\SFE$ 
\be
\SFE\equiv \Msink/M_0 = M_{\star}/M_0,
\label{eq:sfe}
\ee
where $M_0$ is the initial mass of the cloud, while $\Msink$ and $M_{\star}$ are the total mass in sink particles and stars respectively. Note that this metric is normalized by the total cloud mass, so the star formation efficiency of individual self-gravitating subregions can be higher. 
A characteristic time scale of the problem is the initial freefall time $\tff$ of the cloud:
\be
\tff \equiv\sqrt{\frac{3 \pi}{32 G \rho_0}},
\label{eq:tff}
\ee
where $\rho_0$ is the initial cloud density. Combining these two leads to the star formation efficiency per freefall time $\epsff$, a common metric in the literature \citep{Krumholz_Mckee_2005_SF_turbulent_regulated}:
\be
\epsff(t) \equiv  \dot{M}_\mathrm{sink}(t) \tff / M_{\rm gas},
\label{eq:epsff}
\ee
where $M_{\rm gas}$ is the available \emph{initial} gas mass. To better describe the time dependence of these values we also introduce $\tsf$, the number of initial freefall times elapsed since the cloud started star formation:
\be
\tsf(t) \equiv \left(t-t_\mathrm{SF\,starts}\right)/\tff,
\label{eq:tsf}
\ee
where $t_\mathrm{SF\,starts}$ is the time the first star forms in the simulations.

A key prediction of any star formation model is the initial mass spectrum of stars (i.e. the IMF), which we assume to be identical to the mass spectrum of sink particles at the end of the simulation (for caveats see \ref{sec:caveats}). 
To illustrate the evolution of the mass spectrum and to make comparisons easier, it is useful to derive a characteristic mass scale of the stellar population. Note that all such scales are only calculated for the stars above our $\Mcompleteness=0.1\,\msun$ completeness limit. In this work and its appendices we use several different summary statistics of the mass function:
\begin{itemize}
    \item $\Mmedian$: the number-weighted median stellar mass, half of the stars will be more massive than this value. In strongly peaked distributions it provides a good estimate for the peak of the IMF (i.e., the peak of $\dderiv N/\dderiv \log M$). Note that by including only stars above our $\Mcompleteness$ completeness limit, the median stellar mass is not sensitive to mass resolution (see Appendix \ref{app:completeness}).
    \item $\Mmean$: the number-weighted mean stellar mass. Similar to $\Mmedian$ it provides an estimate for the IMF peak.
    \item $\Mmassmedian$: The mass-weighted median stellar mass, half of the stellar mass is in stars more massive than this value. This measures ``where the mass is" in the IMF. This metric is insensitive to low-mass objects and probes the high-mass tail of the IMF, for example in \citetalias{guszejnov_starforge_jets} we found this metric to be strongly affected by the runaway accretion of massive stars that happens if stellar feedback is insufficient.
    \item $\slope$: The effective slope of the IMF between $1-10\,\msun$ that is derived from the mean stellar mass within the same range. Assuming the IMF in that range to be a pure power-law in the form $\dderiv N/\dderiv M \propto M^{-\slope}$, there is a one-to-one mapping between $\slope$ and the mean mass within that range. This metric probes the most well-constrained part of the IMF: observations find this slope to be near-universal in the Local Group with a value of $-2.35\pm 0.25$ \citep{salpeter_slope,kroupa_imf, imf_universality}. 
    \item $\Mmax$: The mass of the most massive star. Since stellar feedback is strongly non-linear with mass, $\Mmax$ is expected to have a major impact on the evolution of the cloud. Note, however, that $\Mmax$ is partially stochastic in our simulations, both due to truly random elements of the simulation and the chaotic nature of the problem \citep{Geen_2018_SF_indeterministic,grudic_starforge_methods}. 
    \item $L/M$: The light-to-mass ratio that is the ratio of the total bolometric luminosity from stars divided by the total stellar mass (in units of $\mathrm{L_\odot}/\msun$). Since stellar luminosity is $L_*\appropto M_*^{7/2}$ this quantity is set by the most massive stars in the simulation and has the same stochasticity issue as $\Mmax$. 
\end{itemize}
 
 Finally, we compare the stellar mass distributions predicted by our simulations with the observed IMF. In simulations where star formation is quenched and the cloud is disrupted we use the post-disruption sink mass distribution as the IMF\footnote{Note that due to the finite resolution of the simulation sink particles may not represent individual stars, but we expect that to be the case above our $\Mcompleteness=0.1\,\msun$ completeness limit.}. In runs that neglect the feedback processes required to quench star formation, we take the sink mass distribution at $\SFE\sim 10\%$ (similar to the values runs with quenched star formation achieve) to be the IMF. We compare these simulated IMFs with the fitting function of \citet{kroupa_imf} that is widely used in the literature\footnote{Note that there are several other fitting functions in the literature (e.g., \citealt{chabrier_imf, De_Marchi_IMF_2010,Parravano_2011_TPL_IMF,Dib2017}), but the differences between them is not significant for the purposes of this paper.}, henceforth referred to as \citetalias{kroupa_imf}. For the mass scales above we calculate the 95\% confidence interval for the individual values at different total stellar masses by varying the IMF with the uncertainty values reported in \citetalias{kroupa_imf} and calculating the various mass scales for each realization. Similarly, for the IMF we take the 95\% confidence intervals in each individual mass bin of the distribution. 
 
  \section{Role of different physics in setting the IMF and star formation history}\label{sec:results_ladder}
  
We carried out a suite of simulations using combinations of the initial conditions and physics from Table \ref{tab:IC_phys}. In this section we concentrate on a set of runs with increasingly complex physics, forming a set of runs we denote as the \myquote{physics ladder} of star formation. The \myquote{ladder} starts from a base model of isothermal MHD and gravity (\textbf{I\_M}, explored in \citetalias{Guszejnov_isoT_MHD}), then adds non-isothermal thermodynamics and protostellar jets (\textbf{C\_M} and \textbf{C\_M\_J} respectively). Note that these have been explored in \citetalias{guszejnov_starforge_jets} with a different thermodynamics treatment that does not directly evolve the radiation fields (ApproxRad, see \citetalias{grudic_starforge_methods}). In our analysis we found no qualitative difference in the star formation history or the IMF between those and the new RHD results. The next step is including stellar radiation (\textbf{C\_M\_J\_R}) and, finally, stellar winds and SNe (\textbf{C\_M\_J\_R\_W}), which we recently explored in a single example simulation in \citet{grudic_starforge_m2e4}. We also include a run with all physical processes except protostellar jets to showcase their importance (\textbf{C\_M\_R\_W}). Due to the computational costs of these simulations, the \textbf{M2e4} clouds were the largest ones the full \myquote{physics ladder} was run on and we use these runs in the subsequent analysis (see Table \ref{tab:IC_phys} for details on the IC). In the following subsections we investigate the effects of each \myquote{rung} of the physics ladder on the star formation history of the cloud (\S\ref{sec:sf_history}) and the stellar mass spectrum (\S\ref{sec:masses_IMF}).
  
\subsection{Star formation history}\label{sec:sf_history}

\begin{figure*}
\begin {center}
\includegraphics[width=0.99\linewidth]{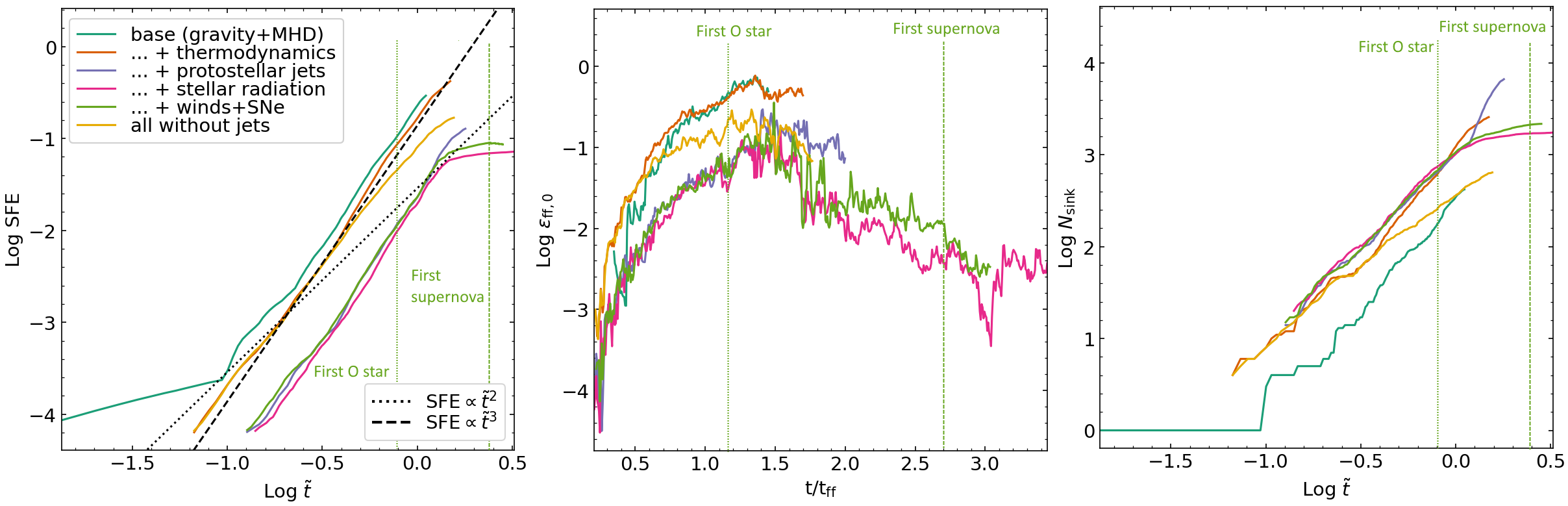}
\vspace{-0.2cm}
\caption{The evolution of the star formation efficiency (left), star formation efficiency per freefall-time $\epsff$ (center) and the number of sink particles $\Nsink$ (right) as function of time for a set of runs with increasingly complex physics in \textbf{M2e4} clouds, corresponding to labels \textbf{I\_M}, \textbf{C\_M}, \textbf{C\_M\_J}, \textbf{C\_M\_J\_R}, \textbf{C\_M\_J\_R\_W}, as well as \textbf{C\_M\_R\_W} (see Table \ref{tab:IC_phys}). The times of the formation of the first O star and the first SN is marked for the full physics (\textbf{C\_M\_J\_R\_W}) case. Note that $t=0$ denotes the start of the simulation, while $\tilde{t}=0$ is set to the start of star formation (see Eq.\ref{eq:tsf}). Without stellar feedback the cloud undergoes runaway star formation. The inclusion of jets suppresses star formation at later times, preventing new stars from forming but still allowing massive stars to accrete. The addition of stellar radiation is required to completely quench star formation. Note that in runs without radiative feedback star formation does not quench before reaching nonphysical values (>10\%), where the runs are terminated.}
\label{fig:sf_history_ladder}
\vspace{-0.5cm}
\end {center}
\end{figure*}

Figure \ref{fig:sf_history_ladder} shows the star formation histories of simulations with identical Sphere initial conditions (\textbf{M2e4}) for different rungs of the \myquote{physics ladder} suite. In all runs the star formation efficiency (SFE) evolves as $\SFE\propto \tilde{t}^3$, similar to the findings for runs without feedback or with protostellar jets only (\citetalias{Guszejnov_isoT_MHD} and \citetalias{guszejnov_starforge_jets} respectively). Note that these results are sensitive to the ICs, see \S\ref{sec:box_vs_sphere} for details. In runs without feedback (\textbf{I\_M},\textbf{C\_M}) star formation continues unregulated. With the addition of protostellar jets (\textbf{C\_M\_J}) star formation is partially suppressed at later times, but continues without cloud disruption. The addition of stellar radiation (\textbf{C\_M\_J\_R}) allows massive stars to greatly influence the cloud evolution, first by blowing away nearby gas (and thus stopping accretion) and by eventually quenching star formation in the cloud. Winds do not qualitatively alter this picture. Meanwhile SNe go off at the end of the full physics run (\textbf{C\_M\_J\_R\_W)}, but it occurs after the cloud is disrupted, too late to meaningfully alter the star formation history of the cloud.

\subsubsection{Cloud disruption}\label{sec:disruption}

\begin{figure*}
\begin {center}
\begin{tabular}{c|c|c}
{\Large\textbf{C\_M}} & {\Large\textbf{C\_M\_J}} & {\Large\textbf{C\_M\_J\_R\_W}}\\
\includegraphics[width=0.32\linewidth]{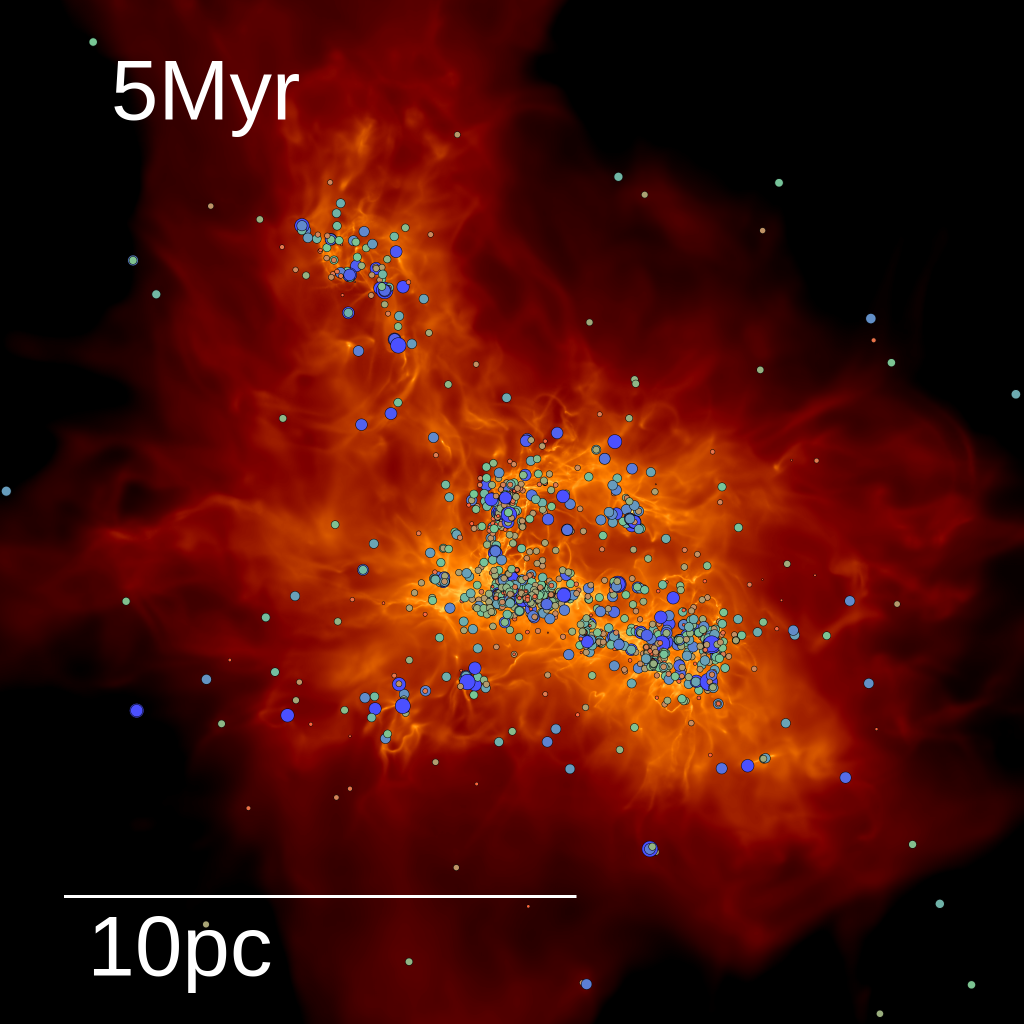} &
\includegraphics[width=0.32\linewidth]{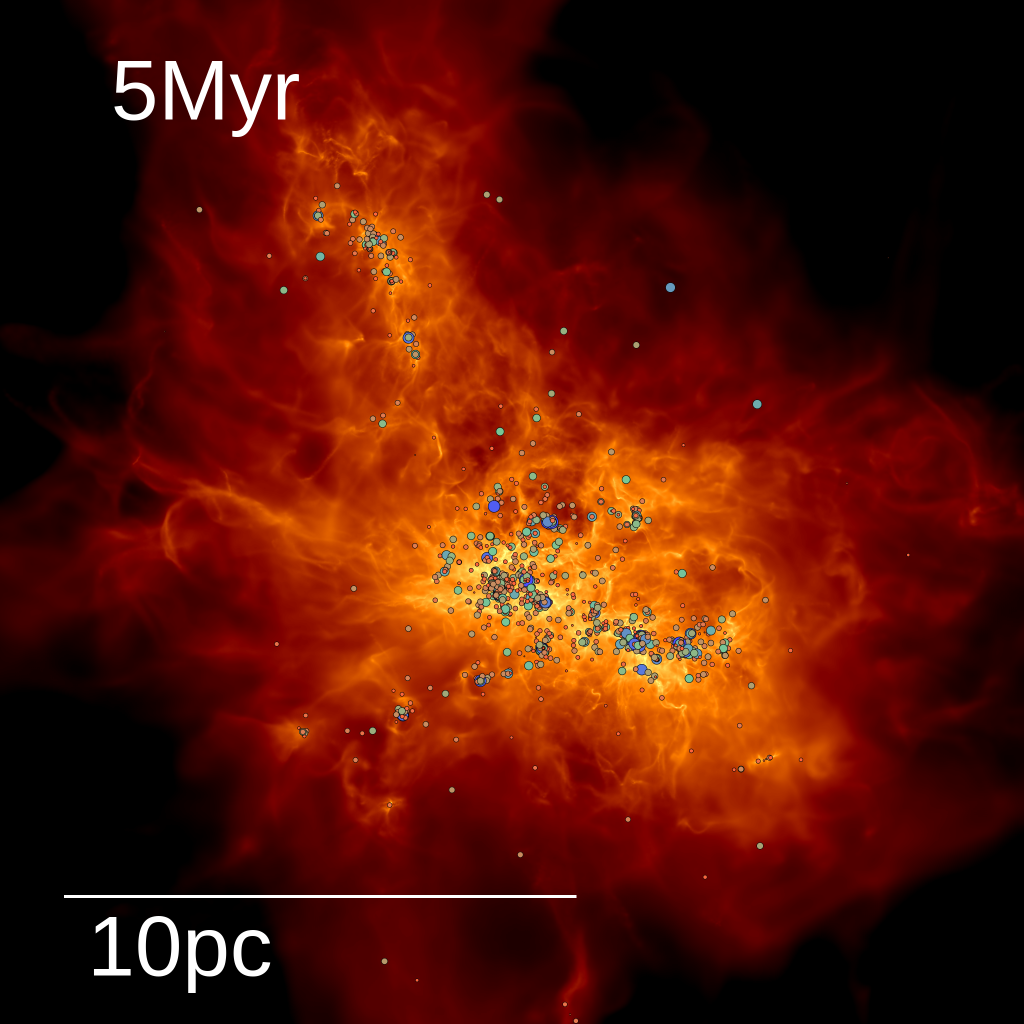} &
\includegraphics[width=0.32\linewidth]{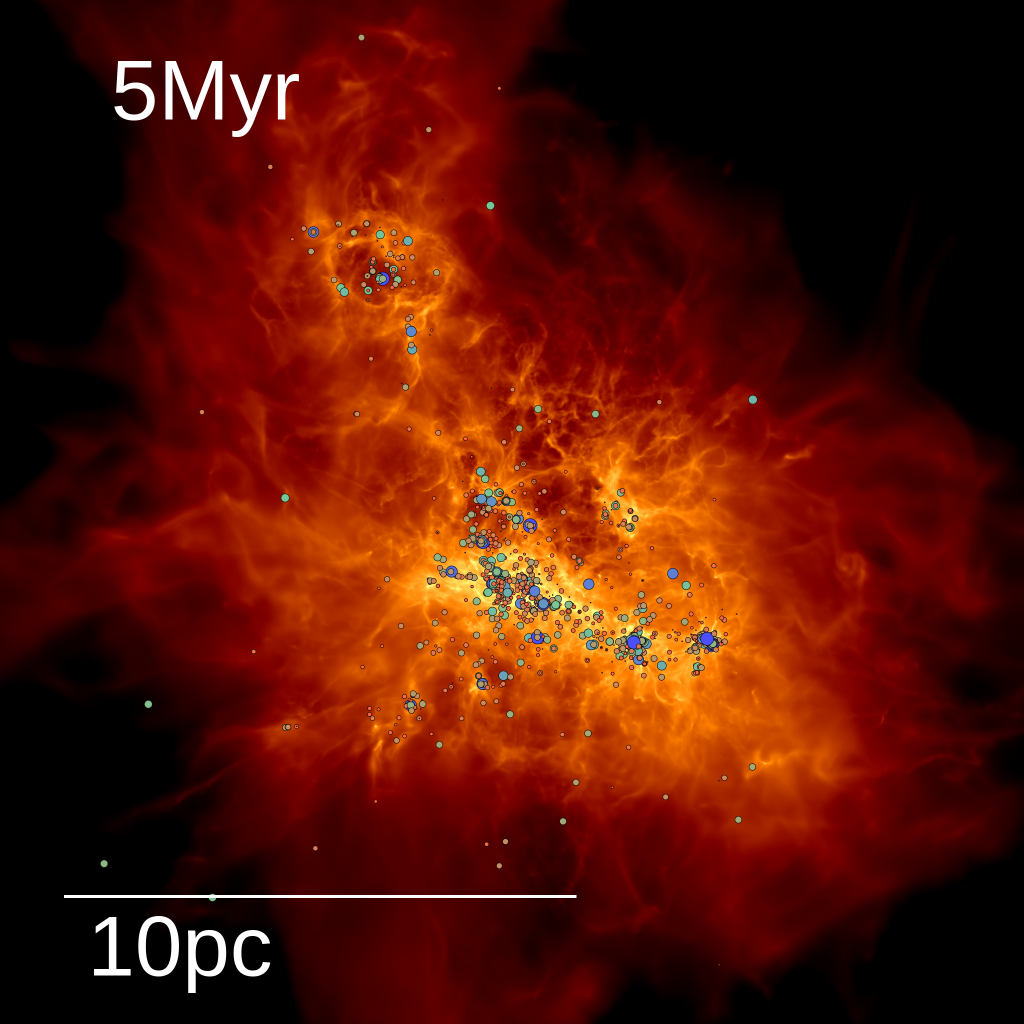}\\
\includegraphics[width=0.32\linewidth]{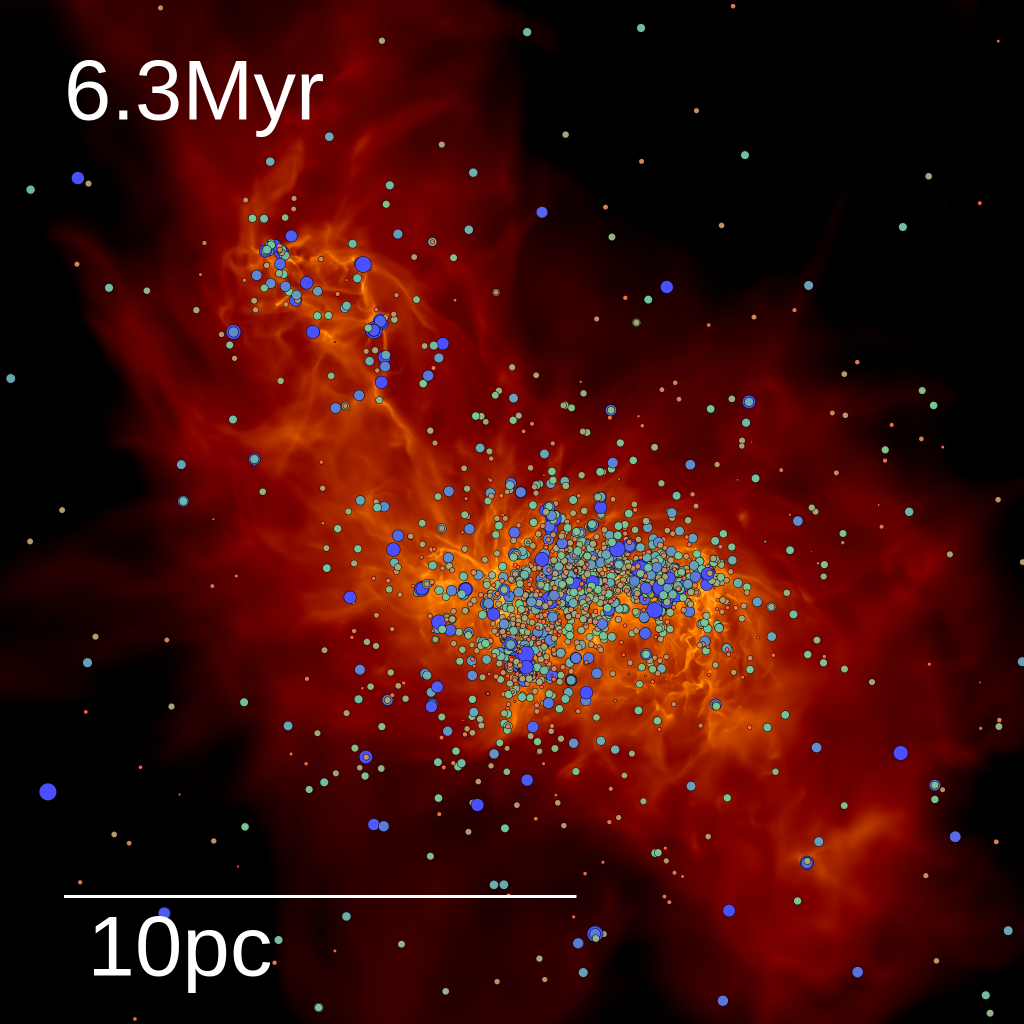} &
\includegraphics[width=0.32\linewidth]{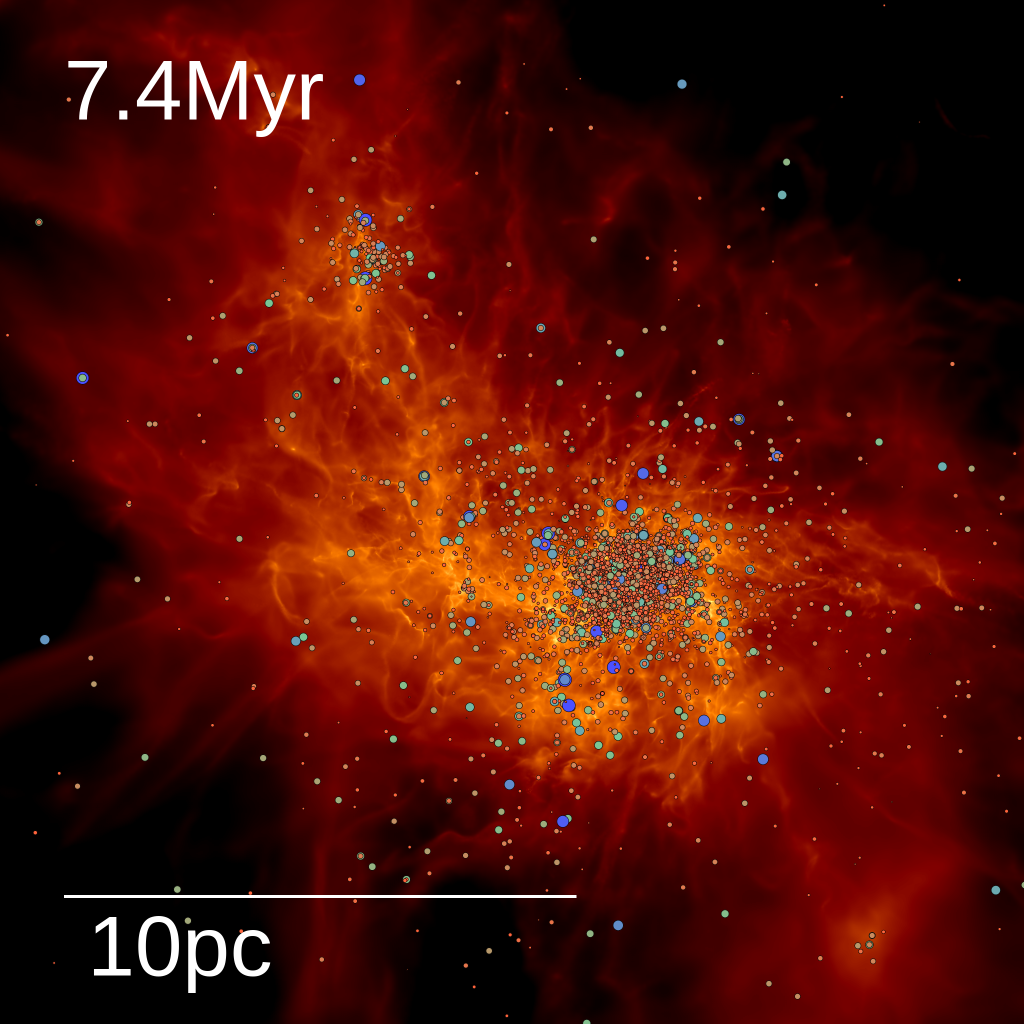} &
\includegraphics[width=0.32\linewidth]{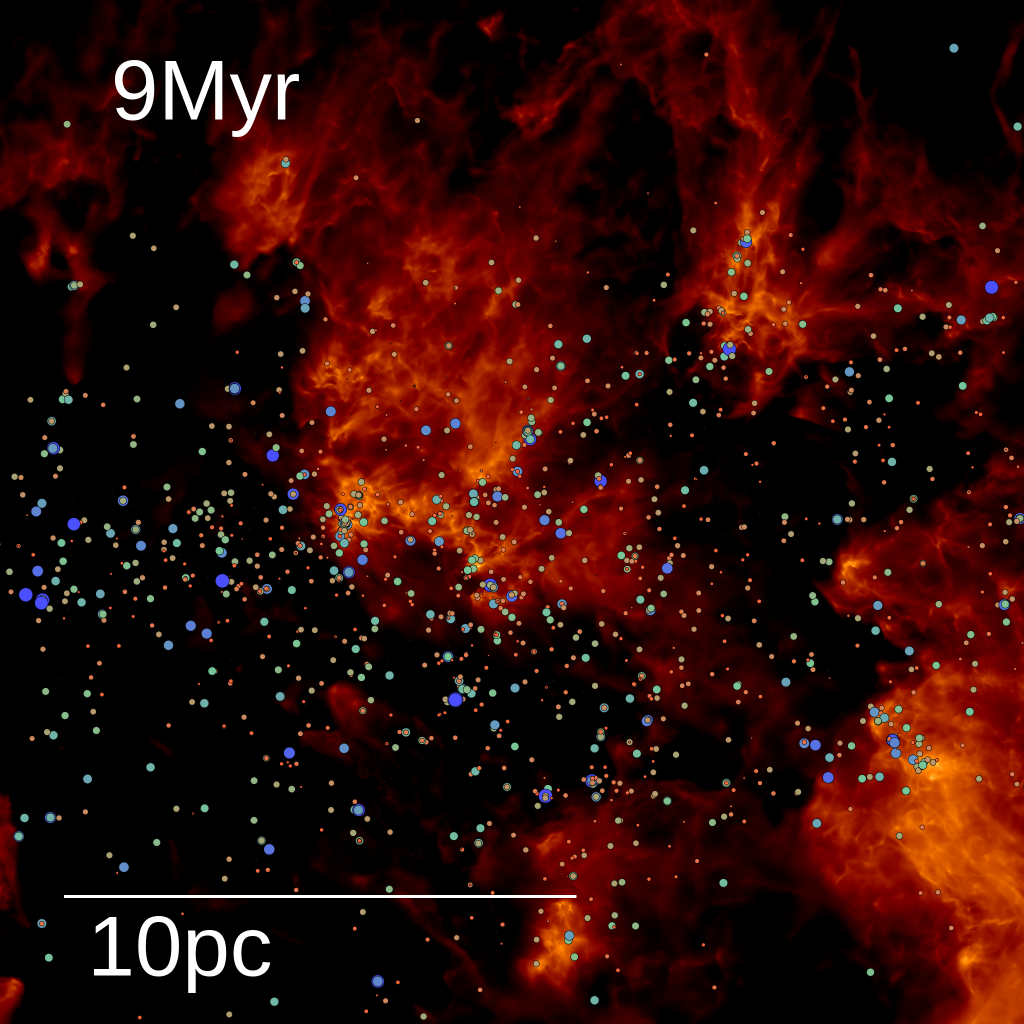}
\end{tabular}
\vspace{-0.4cm}
\caption{Surface density maps at 5 Myr and at the end of simulations for runs with different levels of stellar feedback: no feedback (\textbf{C\_M}, left), protostellar jets only (\textbf{C\_M\_J}, middle) and with jets, radiation, winds and SNe enabled (\textbf{C\_M\_J\_R\_W}, right). The inclusion of additional feedback physics dramatically changes both the sink mass distribution and cloud morphology. Note that the end of the simulations are at different times, as runs that do not experience cloud disruption ((\textbf{C\_M} and (\textbf{C\_M\_J}) are stopped at 10\% SFE.}
\label{fig:disruption_comparison}
\vspace{-0.5cm}
\end {center}
\end{figure*}

One key question of star formation is the regulation of star formation within clouds to achieve a SFE of a few percent \citep{sf_big_problems}. Figures \ref{fig:sf_history_ladder}-\ref{fig:disruption_comparison} show that the simulations from the \myquote{physics ladder} suite attain three qualitatively different endings. Runs without any stellar feedback (\textbf{I\_M}, \textbf{C\_M}) continue to form stars at an accelerated rate as the parent cloud undergoes global gravitational collapse. With the inclusion of protostellar jets (\textbf{C\_M\_J}), global collapse is slowed at high SFE values due to the excess momentum provided by jets to the gas, but star formation continues at a reduced rate. With the inclusion of radiative feedback, massive stars are able to disrupt the cloud and blow away the remaining gas. Note that this blown out gas is still able to form stars, but at a much reduced rate (see late times in Figure \ref{fig:sf_history_ladder}). This late-stage star formation happens late enough in the simulation that SNe from previously formed massive stars can affect it. SNe completely shut down any remaining star formation and blows away all remaining gas.

\begin{figure*}
\begin {center}
\includegraphics[width=0.99\linewidth]{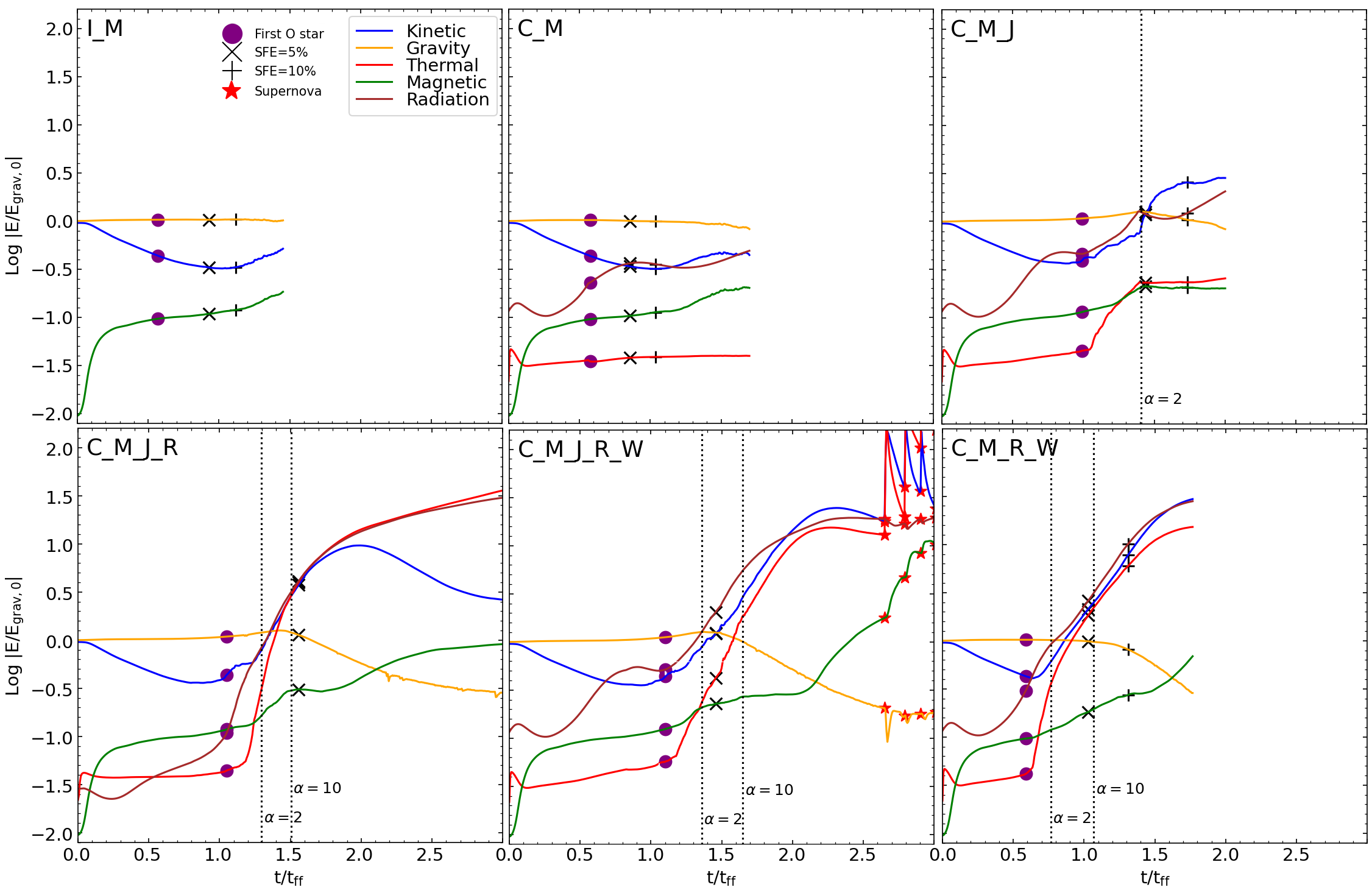}
\vspace{-0.25cm}
\caption{Evolution of cloud energetics
as function of time in the \myquote{physics ladder} suite. Vertical lines mark the times the cloud reaches the boundedness limit ($\alpha=2$) and when the cloud is considered completely disrupted ($\alpha=10$). On the energy evolution trends we mark the appearance of the first main sequence O star (purple circle), the point where a significant portion of the cloud has been turned into stars ($\SFE=5\%, 10\%$, marked with black X and + respectively) and all SNe (red stars). Overall radiative feedback plays a key role in the disruption of the cloud, with the formation of the first O star marking the turning point in the cloud's energy evolution regardless of the presence of jet feedback. Once these massive stars formed kinetic and thermal energy dramatically increases, which leads to the unbinding of the cloud (apparent in the decreasing gravitational binding energy).  
}
\label{fig:disruption_energy_evol}
\vspace{-0.5cm}
\end {center}
\end{figure*}

Figure \ref{fig:disruption_energy_evol} shows the evolution of the cloud energetics within the \myquote{physics ladder} suite. Since we use Sphere ICs (see \S\ref{sec:initial_conditions}) turbulence initially decays in the simulations, until gravitational collapse provides enough kinetic energy to saturate roughly to the virialized values ($\alphaturb\sim 1$). During this time the initial magnetic field is amplified by the turbulent dynamo, increasing the relative magnetic energy to gravitational from $\sim1\%$ to $\sim10\%$ (see \citetalias{Guszejnov_isoT_MHD} and references therein for a discussion). The inclusion of non-isothermal physics does not significantly affect the overall energy evolution of the cloud. Note that the virial parameter of the gas in the late stages of this run does go above the $\alpha=2$ boundedness limit, but the cloud is still bound due to the sink particles (stars). Stellar feedback in the form of protostellar jets dramatically alter the cloud's energetics, as jets entrain nearby gas, creating outflows. The overall effect stellar feedback has on the cloud energetics is a dramatic increase in kinetic and thermal energy, raising the virial parameter to well above the boundedness limit. However, this does not mean that the cloud is fully disrupted and SF is quenched, as a large amount of heated gas remains (see Figure \ref{fig:disruption_comparison}) that can still be accreted by the central cluster, fueling star formation and the growth of existing massive stars. The addition of radiative heating produces initially similar trends as previous runs, but the formation of the first massive, main-sequence O star dramatically affects the evolution of the cloud. Massive stars emit an enormous amount of radiation, outproducing the luminosity of all other stars. Since massive stars form in the dense, central regions of our simulated clouds, a significant portion of their radiation cannot escape, leading to a marked increase in thermal then kinetic energy. This surge in thermal and kinetic energy completely unbinds the cloud relatively quickly, leading to $\alpha>10$ values. Without jets stellar masses are significantly higher, leading to more massive stars and a significantly earlier disruption of the cloud. The addition of stellar winds makes it easier for massive stars to unbind the cloud, but they have no qualitative effect on the simulations in our suite. Finally, supernovae occur only after the cloud has been unbound by radiative feedback. It should be noted that for more massive clouds, that have longer dynamical times, SNe could occur early enough to play a role in cloud disruption.

Note that the disruption time of the cloud is highly sensitive to the time and mass of the first O-type star, which itself is subject to the initial turbulent realization as well as stochastic effects (see \S\ref{sec:realizations}). Thus the effects of different physics or parameters on the cloud disruption time can only be studied in a statistical sense for which we currently have too few simulations.

\subsection{Sink particle Initial Mass Function (IMF)}\label{sec:masses_IMF} 

In this subsection we analyse the effects the different physical processes in the \myquote{physics ladder} have on the stellar mass spectrum. In the simulation sink particles represent stars (or star systems with unresolved stellar companions when the at-formation separation is below the $\dderiv x\sim 30\,\AU$ Jeans-resolution of the simulation). Note that an analysis of a subset of these simulations has already been presented in \citetalias{Guszejnov_isoT_MHD} and \citetalias{guszejnov_starforge_jets}, so here we concentrate on the new results from the simulations that include radiation, winds and SNe feedback processes.

\subsubsection{Evolution of stellar mass scales}\label{sec:mass_scales}

A common issue in numerical simulations is that the low-mass end of the sink mass spectrum is sensitive to numerical resolution and simulations often have a large number of very low-mass objects near their resolution limit. While in most cases these objects represent a vanishingly small fraction of the total sink mass (see \citetalias{Guszejnov_isoT_MHD} for an example and \citealt{guszejnov_isothermal_collapse} for a counterexample), their large number skews the mean and median sink masses. We mitigate these effects by taking the mean and median only for stars more massive than the completeness limit of the simulation. The mass-weighted median mass of sinks $\Mmassmedian$ does not suffer from this effect (see \citealt{krumholz_2012_orion_sims} and \citetalias{Guszejnov_isoT_MHD}), but this choice makes the characteristic mass scale overly sensitive to the most massive sinks, leading to significant variations due to low number statistics (see \S\ref{sec:realizations} for details). Thus, to give a more holistic picture of the evolution of stellar masses, we analyse the evolution of all three mass scales ($\Mmean$, $\Mmedian$, $\Mmassmedian$ and $\Mmax$, see \S\ref{sec:metrics} for definitions).

\begin{figure*}
\begin {center}
\includegraphics[width=0.99\linewidth]{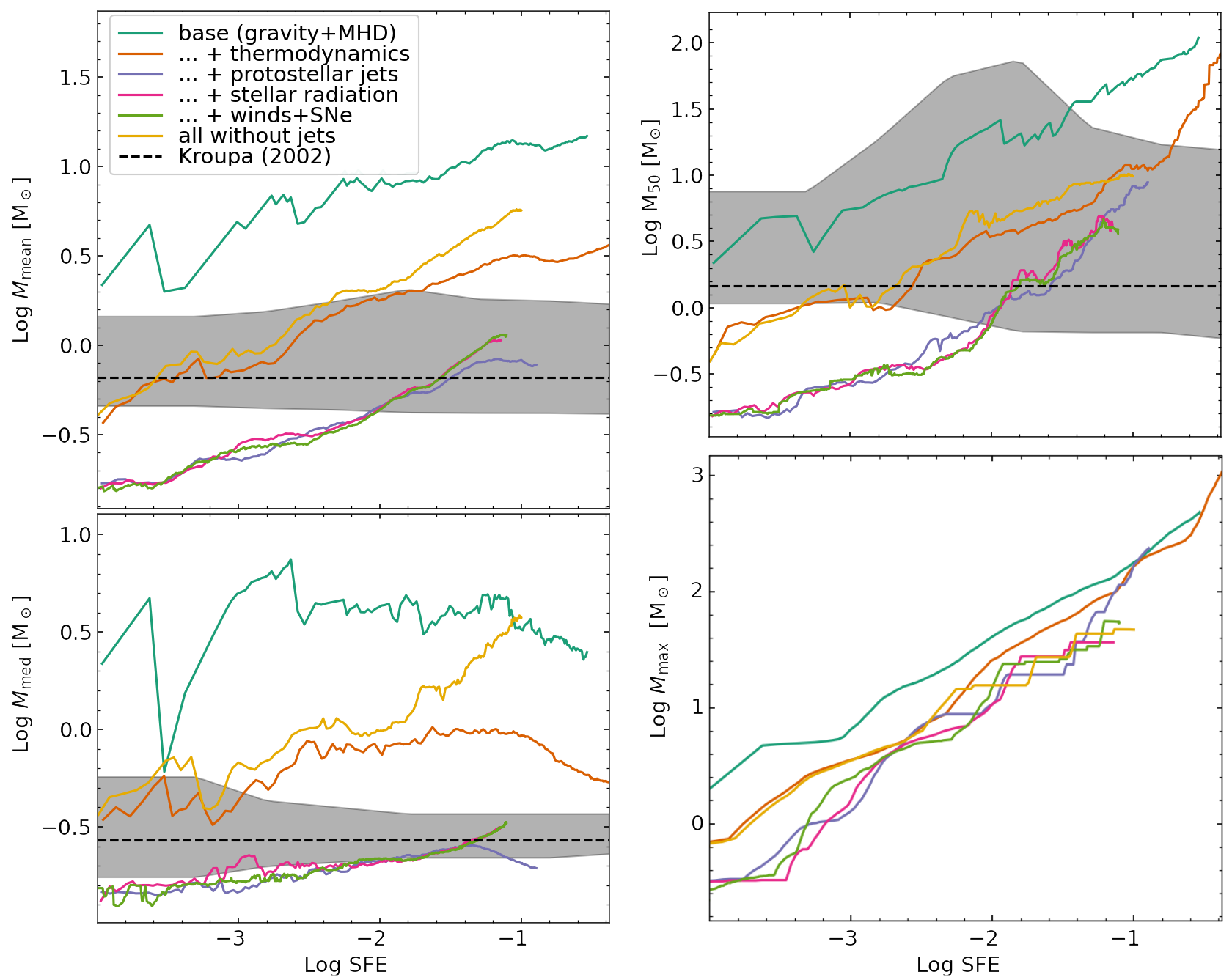}
\vspace{-0.25cm}
\caption{The evolution of the number-weighted mean ($\Mmean = \sum{\Msink}/\Nsink$, top left), number-weighted median (defined such that $\Nsink(M>\Mmedian)=\Nsink/2$, bottom left), mass-weighted median ($\Mmassmedian$, the mass scale above which half the total sink mass resides, top right) and the maximum ($\Mmax$, bottom right ) sink mass as a function of star formation efficiency for the \myquote{physics ladder} suite shown in Figure \ref{fig:sf_history_ladder}. We also show with a shaded region the 95\% confidence interval for these values using the fitting function of \citetalias{kroupa_imf}. These are obtained by constructing a large set of IMFs whose parameters are sampled around the fiducial values with the uncertainties described by \citetalias{kroupa_imf}, then sampling each of these distributions up to the current total stellar mass in the simulation. For these sampled populations we calculate the 95\% confidence interval of the various stellar mass metrics. All simulations are run to $\SFE>10\%$ unless SF is quenched earlier. Protostellar jets reduce sink masses and bring all three mass scales closer to those of the observed IMF, however, $\Mmassmedian$ increases with time, diverging from observations at higher SFE values. }
\label{fig:ladder_mass_scale_evol}
\vspace{-0.5cm}
\end {center}
\end{figure*}

In Figure \ref{fig:ladder_mass_scale_evol} we plot the evolution of these mass scales as a function of star formation efficiency (all simulations are run to $\SFE>10\%$ unless SF is quenched earlier). As in \citetalias{Guszejnov_isoT_MHD} and \citetalias{guszejnov_starforge_jets}, without feedback all mass scales of the sink particles are significantly higher than observed, although the switch to explicit RHD without feedback does allow high-density regions to cool below 10 K, leading to somewhat lower mass stars. Note that this was not the case in \citetalias{guszejnov_starforge_jets}, where radiation was not explicitly evolved and a 10 K floor was enforced (see Table \ref{tab:IC_phys}). Introducing protostellar jets brings low- and intermediate- stellar masses in line with observed values, but SF is only suppressed, not quenched \citepalias{guszejnov_starforge_jets}, while massive stars undergo runaway accretion, which is apparent in both $\Mmassmedian$ and $\Mmax$. With the introduction of radiative feedback the overall mass scales are not affected but massive stars no longer undergo runaway accretion and SF is quenched, leading to a final $\SFE\sim 7\%$.

We find that all four mass scales for all rungs of the \myquote{physics ladder} exhibit an evolution with $\SFE$, thus the final $\SFE$ value where SF is quenched plays an important role in setting the final mass scales of the stellar mass spectrum.  For the clouds in this suite (\textbf{M2e4}, see Table \ref{tab:IC_phys}) stellar winds and SNe did not significantly alter the final SFE value ($\approx 7\%$ in both cases). Overall, for the runs with both jets and radiative feedback we find that
\begin{eqnarray}
\Mmean\propto \SFE^{0.5\pm 0.2}\nonumber\\
\Mmassmedian\propto \SFE^{0.8\pm 0.2}\nonumber\\
\Mmedian\propto \SFE^{0.3\pm 0.2}\nonumber\\
\Mmax\propto \SFE^{0.8\pm 0.2},
\label{eq:mass_scale_SFE}
\end{eqnarray}
where the uncertainty is the mean-squared fitting error. 

\subsubsection{Initial Mass Function}\label{sec:IMF}

\begin{figure*}
\begin {center}
\includegraphics[width=0.47\linewidth]{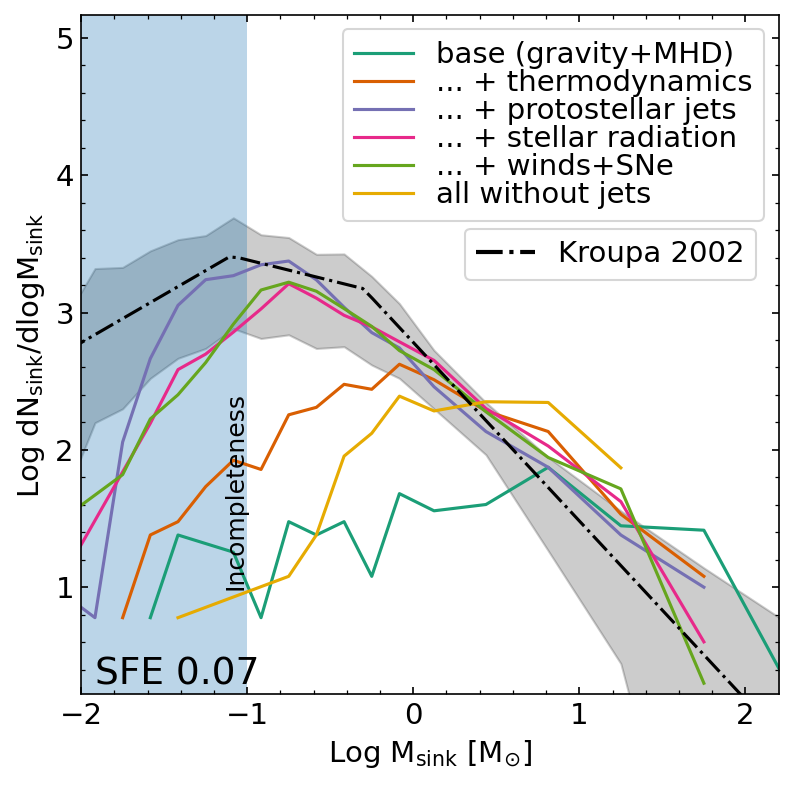}
\includegraphics[width=0.47\linewidth]{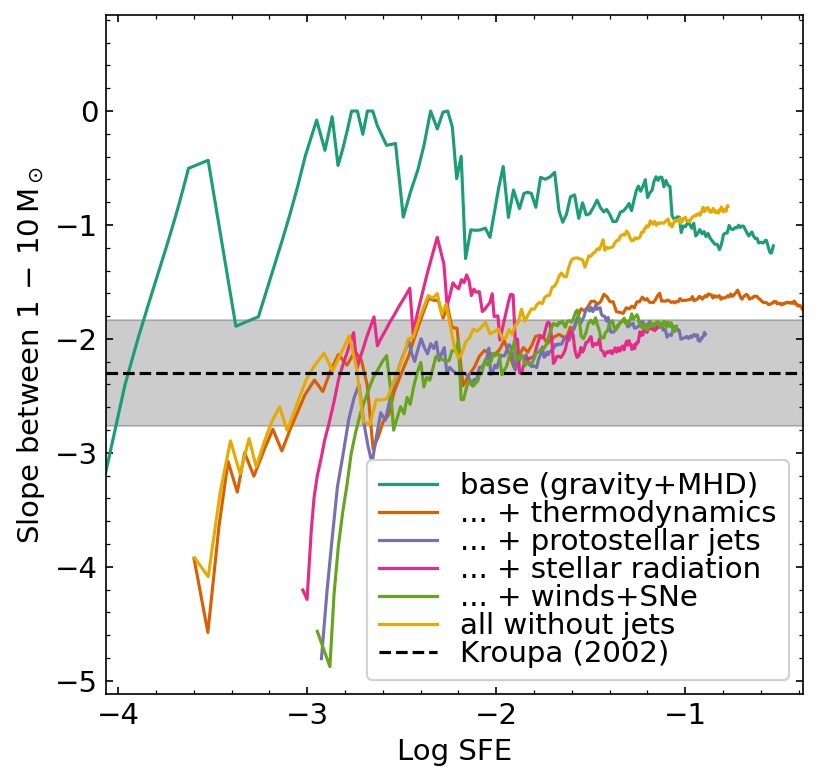}
\vspace{-0.4cm}
\caption{(\textit{Right}) Distribution of sink particle masses measured in each simulation at 5\% star formation efficiency ($\SFE=\sum M_{\mathrm{sink}}/M_{0}$) for the runs shown in Figure \ref{fig:ladder_mass_scale_evol}. We also show the \citetalias{kroupa_imf} fitting function for the IMF with a shaded region illustrating the uncertainties.  (\textit{Left}) The effective slope $\slope$ of the sink mass distribution between $1-10\,\msun$ for the same runs as a function of star formation efficiency along with the 95\% observational confidence interval from \citetalias{kroupa_imf}. This high-mass slope is fairly stable for runs with jet feedback and produce values close to -2. Additional feedback (i.e. radiation, winds) steepen the slope as they are more dominant for massive stars.}
\label{fig:imf_compare}
\vspace{-0.5cm}
\end {center}
\end{figure*}

While the various characteristic mass scales provide some information on the sink (stellar) mass distribution, a holistic view of the sink mass spectrum (IMF) is necessary to understand the effects of each physical process. Figure \ref{fig:imf_compare} shows the mass distribution of sinks at $\SFE=7\%$, corresponding to the final SFE of the clouds with radiative feedback. The base isothermal MHD + gravity model produces an extremely top-heavy IMF with stellar masses a factor 20 higher than observed. The introduction of detailed thermodynamics allows the gas to cool below the 10 K isothermal temperature limit in dense regions, leading to lower stellar masses (see Figure \ref{fig:n_T_Z} later for details). With the addition of protostellar jets stellar masses are dramatically reduced (significantly more than the mass loss from outflows) and the sink mass spectrum takes on a similar shape as the observed IMF. Due to the finite mass resolution of the simulations ($\dderiv m = 10^{-3}\,\msun$) our IMF is incomplete in the brown dwarf regime ($M<0.08\msun$). Nevertheless the IMF peak is essentially identical between all runs with protostellar jets at the same SFE value. Jets are essential in setting the IMF peak; stellar radiation, winds and SNe do not directly affect this mass scale, instead they are mechanisms to quench SF thus setting the final value of the SFE (see \S\ref{sec:disruption}). 

The high-mass end of the IMF shows apparent deviations between models. However, this region is very sensitive to small number statistics (see \S\ref{sec:realizations}), so we use the effective slope between $1-10\,\msun$ as a proxy to compare the high-mass end of the IMF (see \S\ref{sec:metrics}). Figure \ref{fig:imf_compare} shows that all runs with protostellar jets produce an effective slope close to -2. The addition of radiative and wind feedback suppresses the accretion of massive stars and prevent runaway accretion and thus the eventual flattening of the high-mass IMF, they, however, do not change the IMF slope value from -2. Note that the shallow effective slopes for the no-jet runs (\textbf{I\_M}, \textbf{C\_M}) is due to the peak of the distribution being within $1-10\,\msun$, at higher masses they also produce a power-law tail of -2 slope (see \citetalias{Guszejnov_isoT_MHD} and \citealt{guszejnov_scaling_laws} for details).
  
\section{Sensitivity to initial conditions and parameters}\label{sec:results_sensitivity}

In this section we analyze how the results of our full physics runs (\textbf{C\_M\_J\_R\_W} see Table \ref{tab:IC_phys}) depend on changes in the initial conditions. We test for variations in the following initial parameters: the initial cloud surface density ($\Sigma$), virial parameter ($\alphaturb$), magnetization ($\mu$), metallicity ($Z$), as well as the interstellar radiation field (ISRF) and turbulent driving; see Table \ref{tab:var_guide} for specifics. Our aim is to formulate a general expression for how the IMF is affected by variations in different initial parameters and how these variations influence the star formation history of the cloud.

\begin{table*}
    \setlength\tabcolsep{2.0pt} 
	\centering
	\begin{tabular}{ | c | c | c | c | }
	\hline
	Parameter & Default value & Tested variations  & Labels in Table \ref{tab:IC_phys} \\
	\hline
	Initial turbulence & $\alphaturb=2$ (Marginal boundedness) & x0.5, x2 & \textbf{M2e4\_a1}, \textbf{M2e4\_a4}  \\
	\hline
	Surface density & $\Sigma=63\,\msun/\pc^2$ (MW average) & x10, x0.1 & \textbf{M2e4\_R3}, \textbf{M2e4\_R30}  \\
	\hline
	Cloud mass & $M_0=2\times10^4\,\msun$& x0.01, x0.1 & \textbf{M2e2}, \textbf{M2e3}  \\
	\hline
	Mass-to-flux ratio & $\mu=4.2$ (1\% relative magnetic energy) & x0.3, x0.1 & \textbf{M2e4\_mu1.3}, \textbf{M2e4\_mu0.4}  \\
	\hline
	Interstellar Radiation (ISRF) & Solar-circle values  \citep{habing1968,draine_1978_isrf}  & x10, x100 & \textbf{ISRFx10}, \textbf{ISRFx100}  \\
	\hline
	Metallicity & $Z=Z_\odot$ & x0.1, x0.01 & \textbf{Z01}, \textbf{Z001}  \\
	\hline
    \end{tabular}
        \vspace{-0.1cm}
 \caption{List of parameter variations investigated in \S\ref{sec:results_sensitivity} and the relevant IC/physics labels from Table \ref{tab:IC_phys}.}
 \label{tab:var_guide}\vspace{-0.5cm}
\end{table*}

\subsection{Difference between realizations}\label{sec:realizations}

Before analyzing runs with different initial parameters we need to first examine what kind of variations are possible for the \emph{same} initial parameters. We ran three full physics (\textbf{C\_M\_J\_R\_W}) versions of an \textbf{M2e4} cloud that had the same global parameters but used different initial turbulent realizations, i.e., the runs had different initial velocity fields even though the global turbulent parameters (velocity dispersion, power spectrum) were kept identical.

The grey lines in Figure \ref{fig:unimportant_params_IMF} shows that the qualitative star formation history is similar between the runs, however the evolution of the cloud's virial state can show large variations between runs, as it is mainly set by the feedback of massive O stars, making the evolution of $\alpha$  highly sensitive to the formation times and masses of the most massive stars. We note that even though the runs had identical initial, global parameters, the final SFE values varied mildly between 6 and 8 \%. The cloud disruption time (i.e., time when it reaches $\alpha=10$) varies dramatically between runs, between 1.2 and 1.8 freefall times.

Figure \ref{fig:unimportant_params_masses} shows that although stellar masses start out similar in the simulations, variations of up to a factor of 2 can develop in $\Mmedian$ at fixed time. Similarly there are small variations in the effective high-mass slope. The variations in $\Mmax$ cause significant differences in cloud evolution (i.e., cloud disruption time), as cloud disruption is highly sensitive to the formation history of the most massive stars.
  

Table \ref{tab:realizations_summary} shows the summary IMF statistics from \S\ref{sec:metrics} in the 3 simulations that have identical global parameters but different initial turbulent realizations. We compare these with the values we obtain by sampling the \citetalias{kroupa_imf} IMF fitting function between $\Mcompleteness$ and $150\,\solarmass$, while varying the IMF parameters within the uncertainties reported in \citetalias{kroupa_imf}. For all IMF statistics we find that the runs with our fiducial parameters fall within the 95\% confidence intervals we get from sampling \citetalias{kroupa_imf}. In other words, the IMF produced by runs with our fiducial parameters are within observational uncertainties with the \citetalias{kroupa_imf} IMF, although the resulting IMF slope is consistently shallower than the canonical value for all realizations. This means that the canonical -2.3 slope can not be reproduced just by varying the initial turbulent realization. 


\begin{table}
	\centering
	\setlength{\extrarowheight}{4pt}
	\begin{tabular}{ | c | c  | c |}
	\hline
    IMF statistics & Sampling \citetalias{kroupa_imf} IMF&  Simulations \\
    \hline
	$\Mmedian$ [$\solarmass$] & $0.27_{-0.05}^{+0.10}$ & $0.32\pm 0.02$\\
	\hline
	$\Mmean$ [$\solarmass$] & $0.6_{-0.2}^{+1.0}$ & $1.1\pm 0.1$ \\
	\hline
	$\Mmassmedian$ [$\solarmass$] & $1.4_{-0.8}^{+15.5}$ & $4.3\pm 0.5$ \\
	\hline
	$\Mmax$ [$\solarmass$] & $52_{-40}^{+84}$ & $47\pm 7$ \\
	\hline
	$\slope$ & $-2.3_{-0.59}^{+0.54}$ & $-1.93\pm 0.04$ \\
	\hline
	$L/M$ [$\solarluminosity/\solarmass$] & $900_{-800}^{+3900}$ & $900\pm 100$ \\
	\hline
    \end{tabular}
        \vspace{-0.1cm}
 \caption{Summary statistics of runs with identical parameters to the fiducial run but with 3 different turbulent realizations as well as the values expected from a \citetalias{kroupa_imf}. Note that the statistics are all calculated between $\Mcompleteness=0.1\,\solarmass$ and $150\,\solarmass$. For \citetalias{kroupa_imf} the values and their errors are obtained by sampling the IMF at a fixed total stellar mass of $1000\,\solarmass$ while varying its parameters, then taking the median value and the 95\% confidence intervals respectively. For the simulations we simply take the mean and standard variation of the values in the 3 runs. For all statistics the simulated values fall within the confidence intervals we get from random sampling \citetalias{kroupa_imf}.}
 \label{tab:realizations_summary}\vspace{-0.5cm}
\end{table}

\subsection{Initial level of turbulence}\label{sec:sensitivity_alpha}

In a turbulent medium shocks can create self-gravitating overdensities that ultimately form stars. In a globally collapsing medium (i.e. our Sphere ICs, see Figure \ref{fig:M2e4_series}) gravitational compression also triggers star formation. Since the cloud starts without a global infall motion, the initial level of turbulence (set by $\alphaturb$) determines how long the turbulent star formation channel dominates over global collapse, 
as shown in Figure \ref{fig:unimportant_params_IMF}. This is due to higher $\alphaturb$ both enhancing the turbulent SF channel and slowing down the global collapse of the cloud. Previous work has found that in a turbulent medium without global collapse (i.e., with periodic boundary conditions) $\SFE\propto \tilde{t}^2$ (\citealt{federrath_sim_2012,murray_2015_turb_sim,murray_2018_jets} and \citetalias{guszejnov_starforge_jets}), while our previous results showed $\SFE\propto \tilde{t}^3$ for global collapse dominated 
simulations (i.e., in isolated cloud without external turbulent driving, see \citetalias{guszejnov_starforge_jets}). So the net effect a higher $\alphaturb$ is delaying the $\SFE\propto \tilde{t}^3$ regime, effectively lowering $\epsff$. This delays star formation, and leads to an ultimately lower final $\SFE$ value, 10\%, 8\% and 4\% for $\alphaturb$ values of 1,2, and 4 respectively.


Figures \ref{fig:unimportant_params_IMF}-\ref{fig:unimportant_params_masses} shows that although varying the initial $\alphaturb$ turbulent virial parameter significantly affects the star formation evolution of the cloud, the final stellar mass scales and the IMF are insensitive to the initial level of turbulence in the cloud.


\subsection{Cloud surface density}\label{sec:sensitivity_sigma}

Surface density is considered to be a key parameter in determining the star formation history of a cloud \citep[e.g.,][]{fall:2010.sf.eff.vs.surfacedensity, grudic_2016, Li_Vogelsberger_2019_GMC_disrupt}. Figure \ref{fig:unimportant_params_IMF} shows that to be the case for our simulations as well. The average- and high-surface density runs produce similar star formation histories (within the uncertainties of \S\ref{sec:realizations}), however the low surface density run is dramatically different as it only has a single burst of star formation, because feedback from the newly formed stars easily disrupts the low density cloud. $\Sigma$ has a strong effect on the final $\SFE$ values giving 1\%, 8\% and 14\% for $\Sigma$ values of $6.3$, $63$, $630\,\msun/\pc^2$ respectively.


Despite the vastly different star formation history between the low surface density run and the others, the final stellar mass scales and spectra are essentially identical (Figure \ref{fig:unimportant_params_masses}). This is highly desirable in simulations as observed clouds have orders of magnitude variations in surface density \citep{heyer_dame_2015, Miville_Deschenes_2017_MW_GMCs}, while the observed IMF is near universal \citep{imf_universality}.


\subsection{Cloud mass}\label{sec:sensitivity_mass}

\begin{figure*}
\begin {center}
\includegraphics[width=0.99\linewidth]{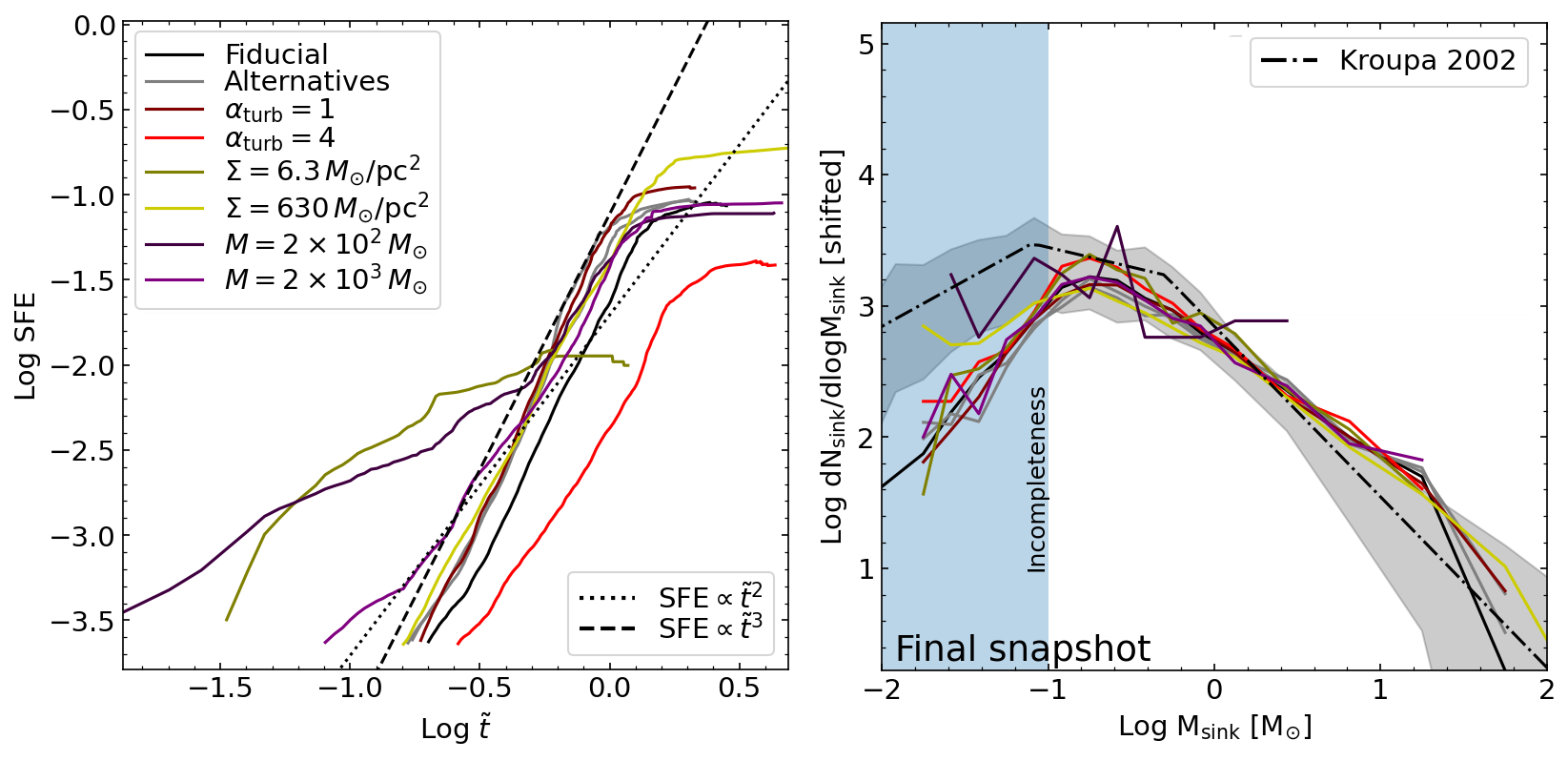}
\vspace{-0.2cm}
\caption{The evolution of the star formation efficiency ($\SFE$, left) and the final sink mass distribution (IMF, right) for runs with identical physics (\textbf{C\_M\_J\_R\_W}). We denote our fiducial \textbf{M2e4} run ($\alphaturb=2$, $\Sigma=63\,\msun/\pc^2$, $M=2\times10^4\,\msun$) with a solid black line. Grey lines show runs with identical parameters but different turbulent realizations. We also show the results of runs with 2x higher and lower virial parameter $\alphaturb$ and 10x higher and lower surface density $\Sigma$ and for lower mass clouds (\textbf{M2e2}, \textbf{M2e3}). With the exception of the low surface density and low mass cases the star formation history is well described by a rising power law that flattens at different final values, ranging between 1-15\%.  Meanwhile the stellar mass distribution (IMF) appears to be nearly invariant to variations in these initial conditions and agrees with the MW IMF within observed uncertainties (shaded area, \citetalias{kroupa_imf}) for stellar masses above $\Mcompleteness$.}
\label{fig:unimportant_params_IMF}
\vspace{-0.5cm}
\end {center}
\end{figure*}

Observed molecular clouds have a large variety of masses from a few thousand to a million $\msun$ \citep{rice2016_mw_gmc_catalogue}, but, due to computational costs, we are only able to probe the $\le 2\times 10^4\,\msun$ range. Figure \ref{fig:unimportant_params_IMF} shows that for these relatively low-mass clouds the star formation history is insensitive to the initial mass for the majority of their lifetime. 
We find that $M_0$ has a negligible effect on the final $\SFE$ values giving 7-8\% values for initial masses of $200$, $2000$ and $2\times10^4\,\msun$ respectively. It should be noted that SF is much more stochastic in the lowest mass cloud due to sampling effects. 

\begin{figure*}
\begin {center}
\includegraphics[width=0.99\linewidth]{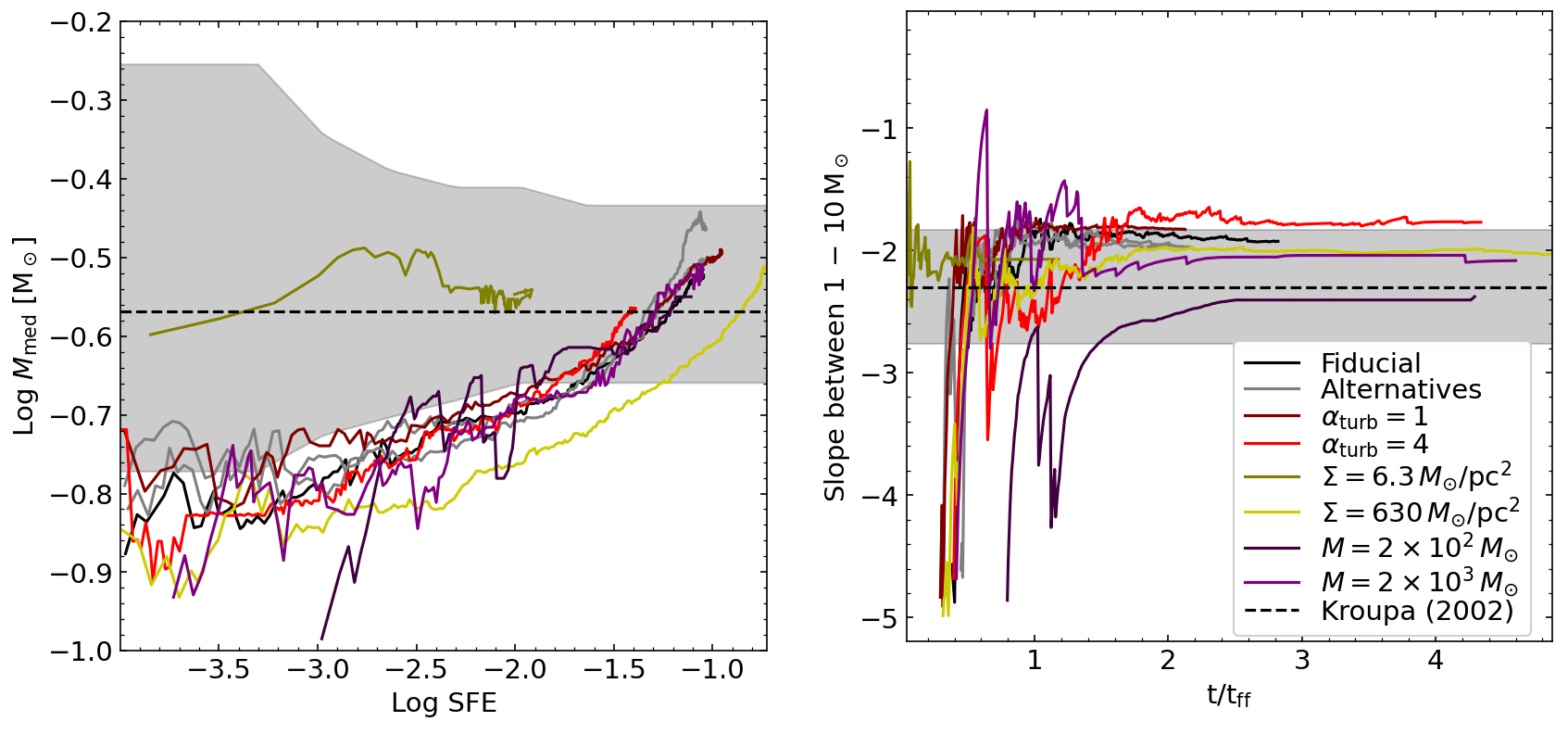}
\vspace{-0.2cm}
\caption{The evolution of the median stellar mass $\Mmedian$ as a function of star formation efficiency $\SFE$ (left) and the evolution of the effective IMF slope $\slope$ (right) for the same runs as in Figure \ref{fig:unimportant_params_IMF} using the same notation. Although $\Mmedian$ varies significantly at fixed $\SFE$, the different runs have different final $\SFE$ values, leading to similar final $\Mmedian$ values for all runs, well within the observational uncertainties. The effective IMF slopes in most runs are shallower than the canonical \citealt{salpeter_slope} value, but are also within observational uncertainties without a significant trend in any of the varied parameters.}
\label{fig:unimportant_params_masses}
\vspace{-0.5cm}
\end {center}
\end{figure*}

Having such similar star formation histories, it is not surprising that the sink mass scales and spectra are also similar (Figures \ref{fig:unimportant_params_IMF}-\ref{fig:unimportant_params_masses}). The lowest cloud mass run (\textbf{M2e2}) does deviate from the higher mass ones, but the difference appears consistent with missing massive stars due to sampling effects. Overall, within the probed mass range the initial cloud mass has no significant effect on any part of the star formation process. Note that this might not be true for more massive clouds, as their longer freefall times (assuming fixed surface density) could allow supernova feedback to affect the cloud evolution (see discussion in \S\ref{sec:discussion}).


\subsection{Cloud magnetization}\label{sec:sensitivity_mu}

\begin{figure*}
\begin {center}
\includegraphics[width=0.99\linewidth]{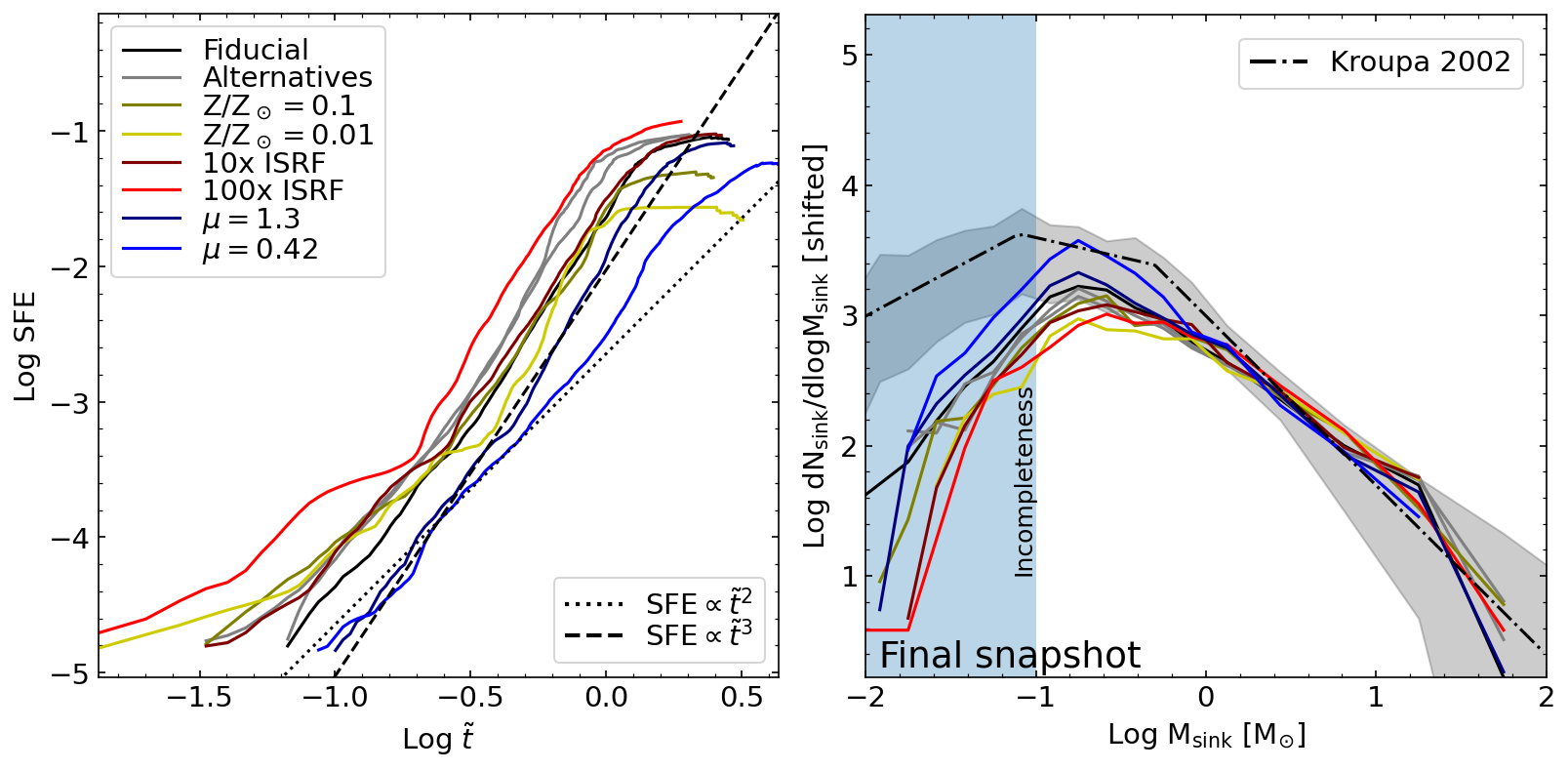}
\vspace{-0.2cm}
\caption{The evolution of the star formation efficiency $\SFE$ (left) and the final sink mass distribution (IMF, right) for runs with identical physics (\textbf{C\_M\_J\_R\_W}). We denote our fiducial \textbf{M2e4} run ($\mu=4.2$, $Z/Z_\odot=1$, $e_\mathrm{ISRF}=e_\mathrm{ISRF,Solar}$) with a solid black line. Grey lines show runs with identical parameters but different turbulent realizations. We also show the results of runs with 3 and 10x higher $\mu$ magnetic fluxes $\alphaturb$ and 10x higher and lower surface density. The star formation history is well described by a rising power law that flattens at different final values, ranging between 3-10\%.  Unlike in Figure \ref{fig:unimportant_params_IMF}, the stellar mass distribution (IMF) appears is somewhat sensitive to these variations, specifically the number of high-mass stars (i.e., high-mass slope of the IMF). However, even these variations are within the observed uncertainties of the MW IMF (shaded area,\citetalias{kroupa_imf}) for stellar masses above $\Mcompleteness$.}
\label{fig:important_params_IMF}
\vspace{-0.5cm}
\end {center}
\end{figure*}

Magnetic fields provide support against collapse \citep{Mouschovias_Spitzer_1976_magnetic_collapse, Shu_1987_star_formation}, and can affect the dynamics of turbulence and feedback in GMCs \citep{maclow:1999.turbulence, krumholz_2007_mhd_hii, offner_2018_mhd_feedback}. Figure \ref{fig:important_params_IMF} shows that in our simulations increasing the strength of the initial magnetic field (corresponding to a decrease in the $\mu$ mass-to-flux ratio) slows down global collapse and significantly reduces the star formation rate of the cloud. Due to the slower star formation rate massive stars in the highly magnetized run have more time to unbind the cloud, resulting in the lower final SFE values of 8\%, 6\% and 4\% for the 4.2, 1.3 and 0.4 mass-to-flux ratio runs respectively.

\begin{figure*}
\begin {center}
\includegraphics[width=0.99\linewidth]{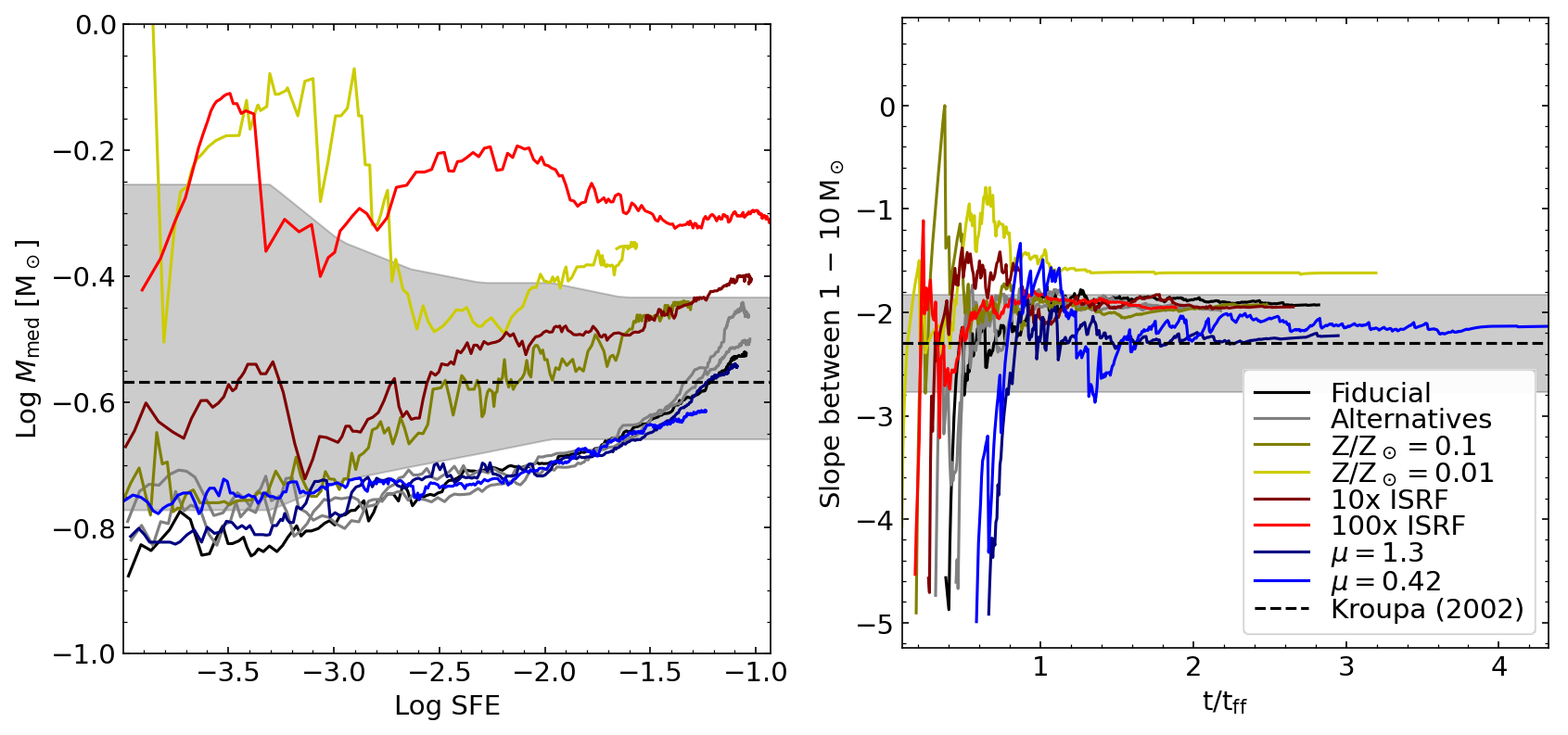}
\vspace{-0.2cm}
\caption{The evolution of the median stellar mass $\Mmedian$ as a function of star formation efficiency $\SFE$ (left) and the evolution of the effective IMF slope $\slope$ (right) for the same runs as in Figure \ref{fig:important_params_IMF} using the same notation. While $\Mmedian$ is insensitive to changes in the initial magnetic field, it is higher at all times if the gas has lower metallicity or a higher background radiation. The effective IMF slopes show mild variations, but only the $z=0.01\,Z_\odot$ run produces slopes shallower than allowed by observational uncertainties.}
\label{fig:important_params_masses}
\vspace{-0.5cm}
\end {center}
\end{figure*}


Higher magnetization leads to a slower star formation and has a significant effect on the IMF. Figures \ref{fig:important_params_IMF}-\ref{fig:important_params_masses} show that high magnetic fields significantly suppresses the formation of massive stars, steepening the high-mass slope to the canonical -2.35 value of \citet{salpeter_slope} above $10\,\solarmass$ in the $\mu=0.4$ case. This most-magnetized run is also the only simulation in this work where jets, radiation and winds are insufficient to unbind the cloud. Star formation is only quenched once the first SN explodes and disrupts the cloud.

\begin{figure}
\begin {center}
\includegraphics[width=0.99\linewidth]{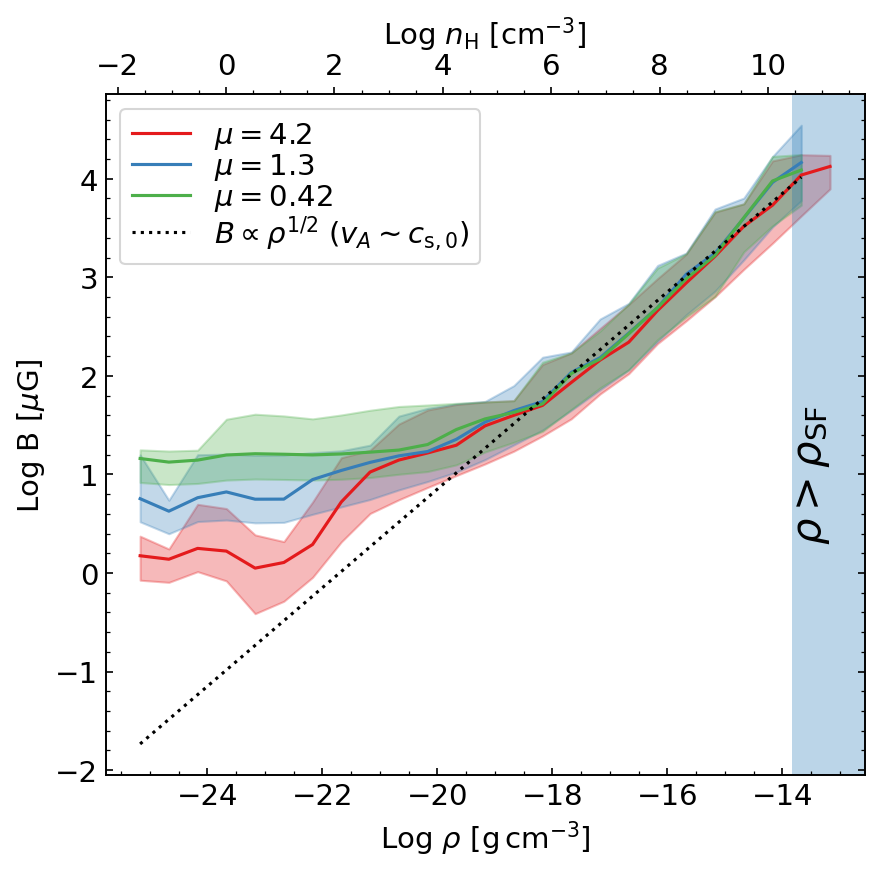}
\vspace{-0.4cm}
\caption{Magnetic field strength as a function of gas density one freefall time after the start of the simulation for runs with different initial normalized mass-to-flux ratios $\mu$. Solid lines show the mass-weighted median values, while shaded regions denote the $\mathrm{1-}\sigma$ (68\%) intervals. To achieve satisfactory statistics we stacked the distribution from 5 snapshots around the target simulation time. We also show $B\propto \rho^{1/2}$ scaling law (in the isothermal regime this corresponds to the $v_A$ Alfvén velocity being an order unity times the thermal sound speed). Note that the simulations have a density threshold $\rho_\mathrm{SF}$ for sink particle formation that we mark with a shaded region, see \citetalias{grudic_starforge_methods} for definition.}
\label{fig:n_B_mu}
\vspace{-0.5cm}
\end {center}
\end{figure}

The relatively minor changes in the IMF peak can be explained by Figure \ref{fig:n_B_mu}, which shows how increasing the initial cloud mass-to-flux ratio only increases the magnetization in low density ($<10^3\,\mathrm{cm}^{-3}$) gas. Despite the different initial conditions, all runs saturate to the same $B\propto \rho^{1/2}$ line, similar to the results of \citet{mocz_2017_core_sim,Wurster_2019_no_magnetic_break_catastrophe} and \citetalias{Guszejnov_isoT_MHD}. This can be understood as the effect of a turbulent dynamo enhancing the magnetic fields and driving the systems towards a {\em common} $B-\rho$ relation at high densities \citep{federrath_2014_dynamo}. This relation roughly corresponds to $v_A(\rho)\sim c_{s,0}$, where $v_A(\rho)$ is the local Alfvén velocity at density $\rho$, while $c_{s,0}$ is the isothermal sound speed, which is the relation we would expect from equipartition. A possible explanation is that the normalization of the $B$-$\rho$ relation is enforced by a local dynamo effect (similar to the global $\alphaB$ saturating in driven boxes, see \citealt{federrath_2011_dynamo}) that is driven by the local gravitational collapse. In numerical experiments, $\beta \sim 1$ is generally achieved for trans- or modestly super-sonic turbulence \citep{stone_1998_mhd_dynamo}, which was indeed found on all scales in individual collapsed cores by \citet{mocz_2017_core_sim}. The fact that this $B-\rho$ relation at high densities is independent from the initial magnetization explains why there are no differences in the mass scale of low-mass stars (which dominate $\Mmedian$), as they accrete most of their material from their natal core that has $n>10^5\mathrm{cm}^{-3}$. Massive stars in these simulations accrete material from scales  much larger than their natal cores \citep{grudic_starforge_m2e4}, so changes in magnetic support at lower densities can affect their accretion flow \citep{lee:2014.magnetized.bh.accretion} .

\subsection{Cloud metallicity}\label{sec:sensitivity_Z}

The thermodynamic behavior of the gas strongly depends on its elemental and dust abundances (metallicity), which is expected to strongly affect the mass scale of stars \citep{larson2005, Sharda_Krumholz_2021_IMF_bottom_heavy_Z}. Also, metal line cooling becomes weaker at low $Z$, which increases the temperature of HII regions \citep{osterbrock_1989_ism}, making feedback more mechanically efficient and reducing the SFE \citep{he_2019_gmc_fb}. Figure \ref{fig:important_params_IMF} shows that in our simulations with different metallicity values the star formation histories are qualitatively similar, i.e., an initial $\propto \tsf^2$ phase followed by $\propto \tsf^3$, but the transition is delayed at low metallicity. Abundances also affect the final star formation efficiency, giving 8, 5 and 3\% for metallicity ($Z/Z_\odot$) values of $1$, $0.1$ and $0.01$ respectively.


Despite the similar (but delayed) star formation rates, the stellar mass scales are significantly different. Figure \ref{fig:important_params_masses} shows that lower metallicities lead to increased stellar masses and the final stellar mass spectrum (IMF) is consistently more top heavy (i.e., has shallower slope) for lower metallicity values. This is due to the less efficient cooling of the gas with the absence of metals.

\begin{figure}
\begin {center}
\includegraphics[width=0.99\linewidth]{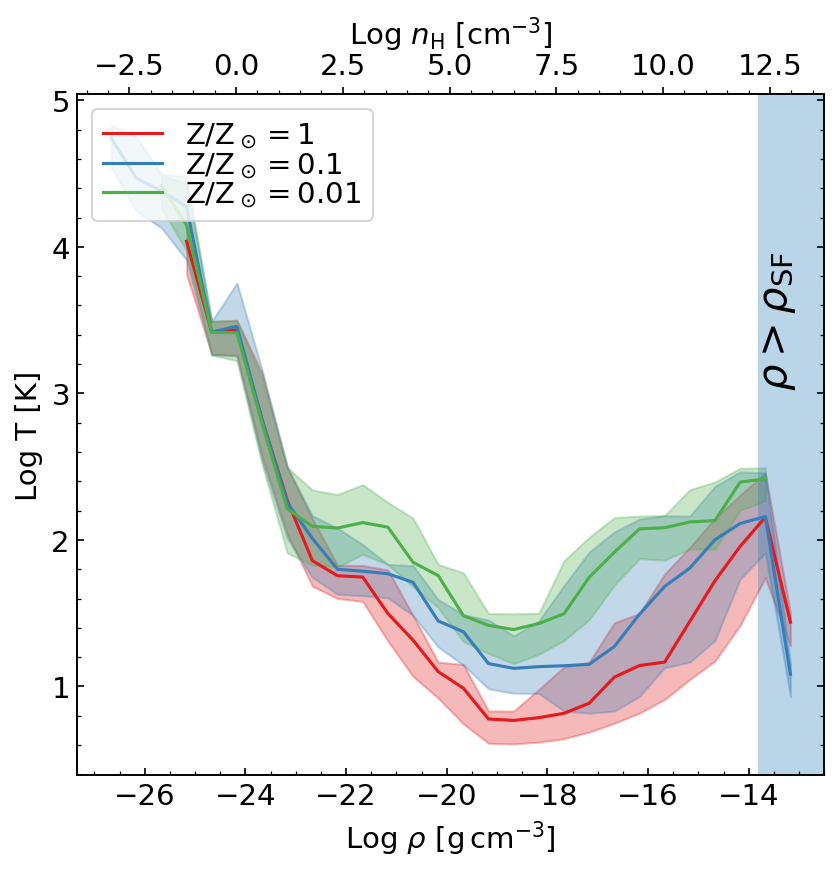}
\vspace{-0.4cm}
\caption{Median temperature of the gas at different densities one freefall time after the start of the simulation for different initial $Z$ metallicity values. Solid lines show the mass-weighted median values, while shaded regions denote the $\mathrm{1-}\sigma$ (68\%) intervals. To achieve satisfactory statistics we stacked the distribution from 5 snapshots around the target simulation time. Note that the simulations have a  $\rho_\mathrm{SF}$ density threshold for sink particle formation that we mark with a shaded region. 
}
\label{fig:n_T_Z}
\vspace{-0.5cm}
\end {center}
\end{figure}

Figure \ref{fig:n_T_Z} shows that lowering the cloud metallicity increases the gas temperature at densities above $\sim 10^3\,\mathrm{cm}^{-3}$. This can be understood as lowering metallicity suppresses molecular line cooling, the dominant cooling channel at densities $\sim 10^3\,\mathrm{cm}^{-3}$, as well as reducing dust density, which in turn reduces dust cooling, the dominant cooling channel above $\sim 10^5\,\mathrm{cm}^{-3}$. We find the resulting temperature to roughly follow
\be
T \propto Z^{-1/4},
\label{eq:Z_to_T_conv}
\ee
which is roughly consistent with what one would expect from blackbody radiation from dust in an optically thin medium that is in equilibrium with the ISRF (i.e., $e_\mathrm{ISRF}\times \mathrm{const.} = j_\mathrm{cool}\propto Z T^4$). 

\subsection{Interstellar Radiation Field}\label{sec:sensitivity_ISRF}

The interstellar radiation field (ISRF) is the only external heating source in the simulation. The ISRF sets the temperature of the gas in the absence of other sources (i.e. before stars form). Increasing the ISRF increases thermal support in the cloud but has little effect on the speed of global collapse (see Figure \ref{fig:important_params_IMF}). Star formation rates between the fiducial and the 10 times higher ISRF runs are similar, but the highest, 100 times larger ISRF run has significantly higher star formation rates. This, in turn, means mildly higher final SFE values with higher ISRF: 8\%, 10\% and 11\% for Solar-Circle, 10 times higher and 100 times higher values respectively. 


Figure \ref{fig:important_params_masses} shows that increasing the ISRF leads to higher stellar masses, due to higher gas temperatures, which lead to an increase in most characteristic mass scales (e.g., $\MJeans$, $\MBE$). This effectively shifts the IMF to higher masses, leading to a shallower effective high-mass slope $\slope$, even though the actual shape (e.g., high-mass slope) of the IMF is largely unchanged (see Figure \ref{fig:important_params_IMF}).

\begin{figure}
\begin {center}
\includegraphics[width=0.99\linewidth]{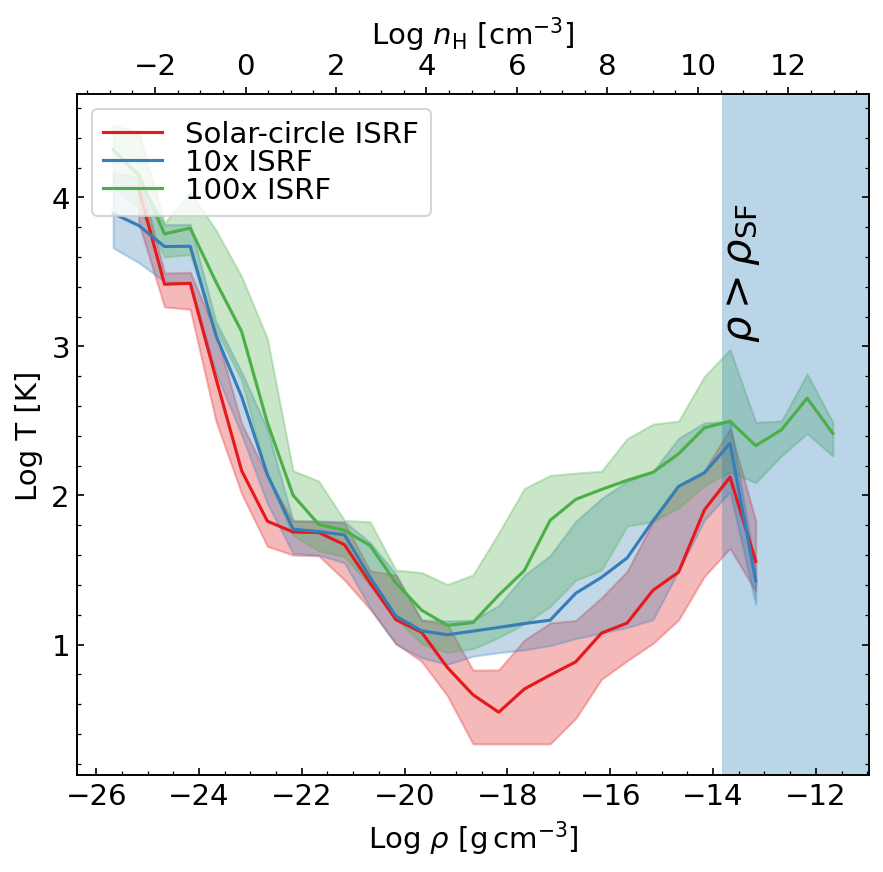}
\vspace{-0.4cm}
\caption{Median temperature of the gas at different densities one freefall time after the start of the simulation for runs with interstellar radiation fields (ISRF) of different strengths. Solid lines show the mass-weighted median values, while shaded regions denote the $\mathrm{1-}\sigma$ (68\%) intervals. To achieve satisfactory statistics we stacked the distribution from 5 snapshots around the target simulation time. Note that the simulations have a density threshold $\rho_\mathrm{SF}$ for sink particle formation that we mark with a shaded region.}
\label{fig:n_T_ISRF}
\vspace{-0.5cm}
\end {center}
\end{figure}

Figure \ref{fig:n_T_ISRF} shows that increasing the ISRF increases the gas temperature at densities above $\sim 10^5\,\mathrm{cm}^{-3}$. This can be understood as increasing the ISRF directly increases the heating radiation with which dust grains interacts, effectively raising the dust temperature. This, in turn, raises the gas temperature above the density where gas becomes strongly coupled to dust ($n> 10^5\,\mathrm{cm}^{-3}$). We find the resulting gas temperature roughly follows
\be
T \propto e_\mathrm{ISRF}^{1/4},
\label{eq:ISRF_to_T_conv}
\ee
similar to Equation \ref{eq:Z_to_T_conv} and is roughly consistent with what one would expect from blackbody radiation from dust in an optically thin medium that is in equilibrium with the ISRF. 

\subsection{Role of turbulent driving}\label{sec:box_vs_sphere}

In our previous works \citepalias{Guszejnov_isoT_MHD,guszejnov_starforge_jets} we found that the star formation history of a cloud significantly differed between the globally collapsing Sphere ICs (similar to \citealt{bate_2009_rad_importance}) and driven, periodic Box ICs (similar to \citealt{Federrath_2014_jets,Cunningham_2018_feedback}). Here we investigate the effects of turbulent driving and the periodic boundary condition by comparing our default full physics (\textbf{C\_M\_J\_R\_W}) Sphere run with two Box runs, one with continuously driven turbulence and one where turbulence is allowed to decay after SF starts. 

\begin{figure*}
\begin {center}
\includegraphics[width=0.99\linewidth]{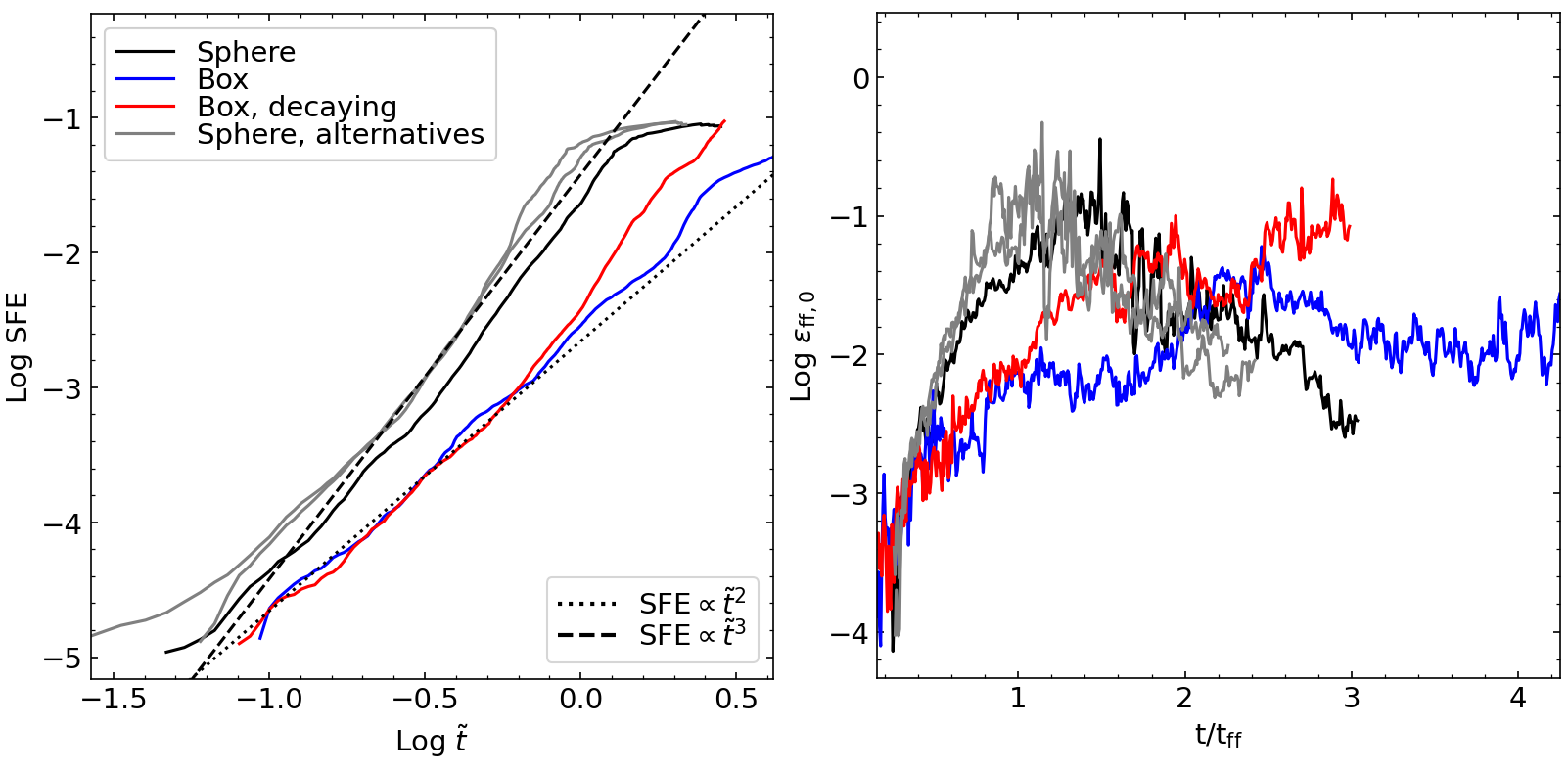}
\vspace{-0.2cm}
\caption{The evolution of the star formation efficiency (SFE) and the star formation rate per freefall-time ($\epsff$) as a function of time for \textbf{C\_M\_J\_R\_W} runs using Sphere IC, Box IC and Box IC with decaying turbulence. Similar to Figure \ref{fig:unimportant_params_IMF} we denote the results for different turbulent realizations for the Sphere run with grey lines. Note that due to its periodic geometry Box runs do not experience cloud disruption, the runs are terminated once the simulated volume is filled with unphysical levels of radiation. In our fiducial Sphere runs SF progresses as $\SFE\propto \tsf^3$, however in the driven Box run it only rises as $\SFE\propto \tsf^2$. We attribute this to the external driving and weaker gravitational focusing, as the decaying Box run transitions between the two regimes as its turbulence decays. Note that in the driven Box case the star formation rate $\epsff$ is roughly steady while in the other cases it varies orders of magnitude over the cloud lifetime.}
\label{fig:m2e4_driving_SF_history}
\vspace{-0.5cm}
\end {center}
\end{figure*}

Figure \ref{fig:m2e4_driving_SF_history} shows that the Sphere and Box runs exhibit the same star formation scaling that was found in the literature, $\SFE\propto\tilde{t}^3$ and $\SFE\propto\tilde{t}^2$ respectively (see e.g., \citetalias{guszejnov_starforge_jets} and \citealt{murray_2018_jets}). The Box run without driving follows an intermediate behavior, initially following a $\SFE\propto\tilde{t}^2$ trend, then switching over to $\SFE\propto\tilde{t}^3$ as turbulence decays and global collapse starts, similar to the high $\alphaturb$ Sphere runs in Figure \ref{fig:unimportant_params_IMF}. Note that due to the periodic boundary conditions in the Box runs neither gas nor radiation can escape the cloud. This means that Box runs do not experience cloud disruption, instead radiation rises in them to unphysical levels as star formation progresses, thus there is no physically meaningful \myquote{final SFE} value for Box runs.
  

\begin{figure*}
\begin {center}
\includegraphics[width=0.99\linewidth]{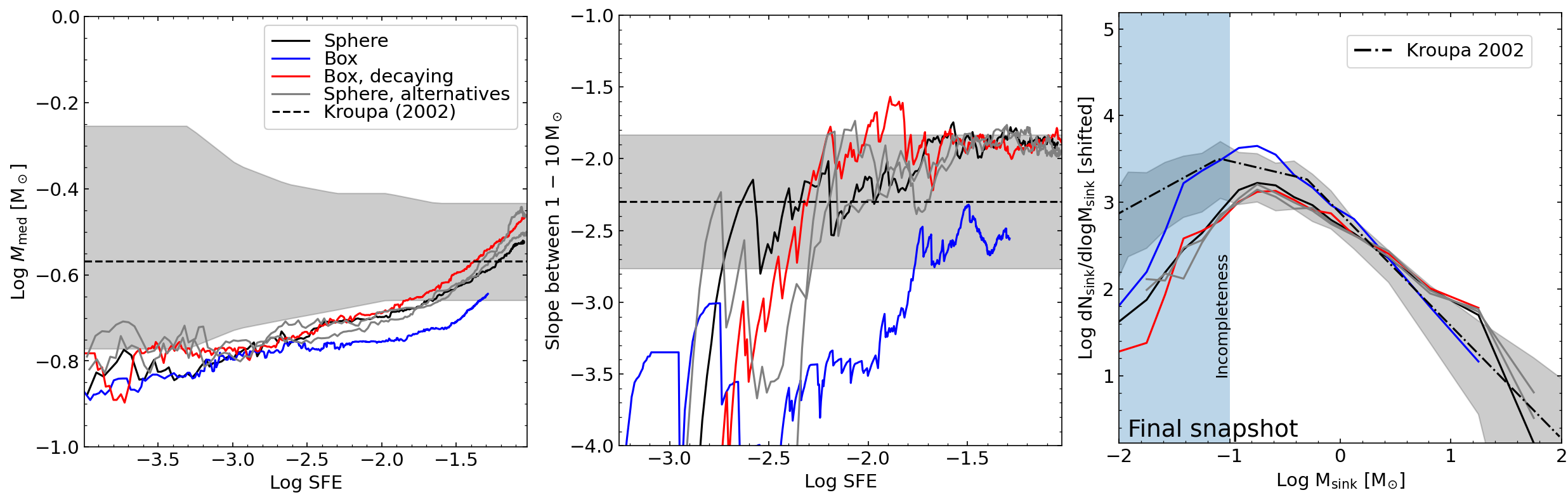}
\vspace{-0.2cm}
\caption{The evolution of the number-weighted median ($\Mmedian$, left) stellar mass, the effective IMF slope ($\slope$, middle) as a function of star formation efficiency $\SFE$ and the final sink mass distribution (IMF, right). The driven Box run exhibits lower stellar masses and a significantly steeper high-mass slope.}
\label{fig:m2e4_driving_IMF}
\vspace{-0.5cm}
\end {center}
\end{figure*}

Figures \ref{fig:m2e4_driving_SF_history}-\ref{fig:m2e4_driving_IMF} show that there is a significant discrepancy between the stellar spectra of the Sphere and Box runs, with Sphere runs producing about a factor of 2 higher stellar masses. Thus the effective slope $\slope$ is steeper for the Box runs than observed but still within the limits of sampling uncertainty. Stellar masses in the Box run without turbulent driving start out similar to the driven case, but quickly switch over to the same track as the Sphere run. Note that observed molecular clouds experience both global collapse and external driving \citep{heyer_dame_2015}, so the behavior of a realistic cloud would likely lie between the tracks of the Sphere and driven Box runs. So we conclude that an accurate modeling of external driving of turbulence in clouds is necessary for any simulation to reproduce the observed IMF (see recent work by \citealt{Lane_2022_TurbSphere} for such a model).

\subsection{Scaling relations}\label{sec:variation_fits}

From the previous results we can formulate a general expression for the dependence of the IMF and the star formation history of the cloud as a function of initial global parameters. Specifically, we attempt to formulate scaling relations for the $\Mmedian$ median stellar mass (above $0.1\,\msun$), the $\slope$ effective IMF slope and the terminal $\SFE$ of the cloud. This is done by carrying out least-squares fitting for the dependence of each individual global parameter while marginalized over the rest. We estimate the error in $\Mmedian$ and $\slope$ by bootstrapping: we resample the sink mass distribution at fixed total sink mass and calculate the 66\% (1-$\sigma$) confidence intervals. We obtain the scaling relations shown in Table \ref{tab:scalings}. We can simplify this scaling relation by ignoring all low and insignificant exponents, i.e., all exponents $\gamma$ with fitting error $\sigma$ where either $|\gamma|<2\sigma$ or $|\gamma|<0.1+\sigma$. 
We find that
\begin{equation}
\Mmedian \sim \mathrm{const.},
\label{eq:fit_mmedian}
\end{equation}
in other words the median stellar mass is insensitive to all varied parameters. For $\slope$ we find that it varies as
\begin{equation}
e^{\slope} \propto   \mu^{0.28\pm 0.16}  Z^{-0.13\pm 0.10}
\label{eq:fit_slope}
\end{equation}
while for the terminal star formation efficiency we get
\begin{equation}
\SFE_\mathrm{final} \propto  \alphaturb^{-0.72\pm 0.24}\, \Sigma^{0.59\pm 0.17}\, \mu^{0.19\pm 0.06}\, Z^{0.26\pm 0.04}.
\label{eq:fit_SFE}
\end{equation}


\begin{table*}
    \setlength\tabcolsep{2.0pt} 
	\centering
	\begin{tabular}{ | c | c | c | c | }
	\hline
	Parameter & Final $\SFE$ exponent &  $\Mmedian$ exponent  & $e^{\slope}$ exponent \\
	\hline
	Initial turbulence ($\alphaturb$) & $-0.72\pm 0.24$ & $-0.12\pm 0.03$ & $0.06\pm 0.27$  \\
	\hline
	Surface density ($\Sigma$) & $0.59\pm 0.17$ & $0.01\pm 0.01$ & $-0.07\pm 0.10$  \\
	\hline
	Cloud mass ($M_0$) & $0.03\pm 0.02$ & $0.0\pm 0.03$ & $0.23\pm 0.19$  \\
	\hline
	Mass-to-flux ratio ($\mu$) & $0.19\pm 0.06$ & $0.10\pm 0.01$ & $0.28\pm 0.16$  \\
	\hline
	Interstellar Radiation (ISRF) & $0.06\pm 0.02$  & $0.11\pm 0.01$ & $-0.03\pm 0.06$  \\
	\hline
	Metallicity ($Z$) & $0.26\pm 0.04$ & $-0.08\pm 0.02$ & $-0.13\pm 0.10$  \\
	\hline
    \end{tabular}
        \vspace{-0.1cm}
 \caption{List of exponents obtained by least-squares fitting for the final SFE value, median stellar mass $\Mmedian$ and the $\slope$ effective high-mass slope IMF in \S\ref{sec:results_sensitivity}.}
 \label{tab:scalings}\vspace{-0.5cm}
\end{table*}

\subsection{Variations in the light-to mass ratio}\label{sec:LM_var}

\begin{figure*}
\begin {center}
\includegraphics[width=0.48\linewidth]{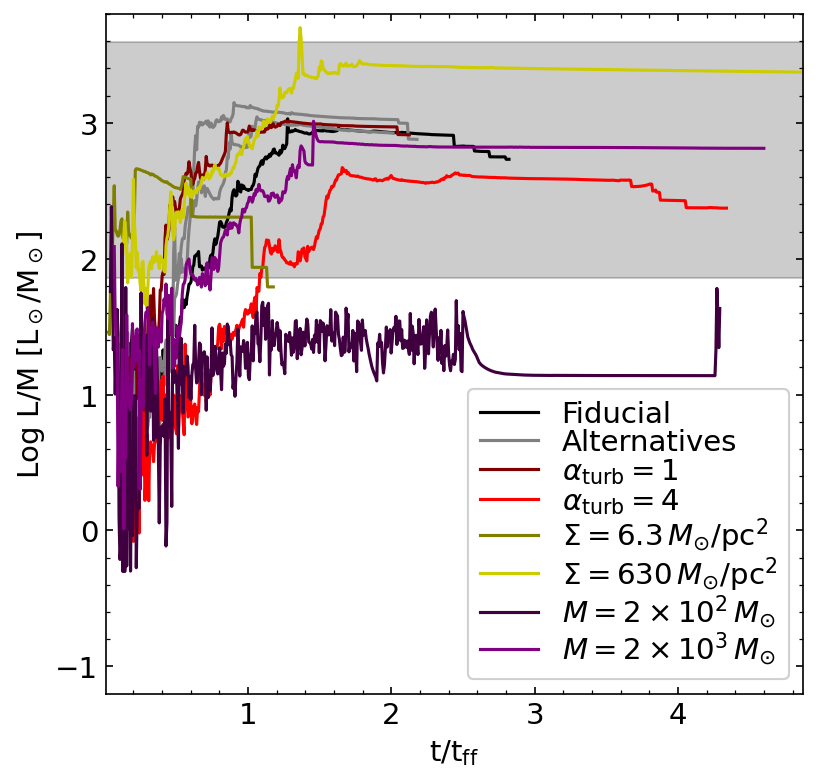}
\includegraphics[width=0.45\linewidth]{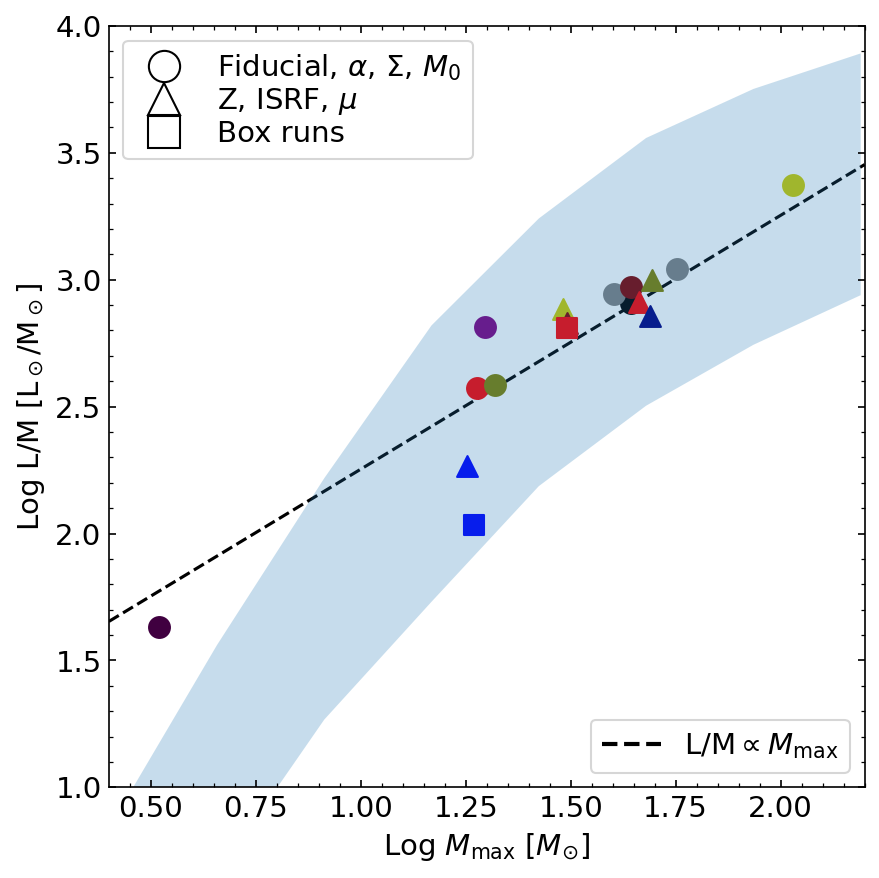}
\vspace{-0.2cm}
\caption{(\textit{Left}) The evolution of the light-to-mass ratio as a function of time for the same set of runs as in Figure \ref{fig:unimportant_params_IMF}. The shaded region denotes the $\mathrm{1-}\sigma$ (68\%) intervals for $L/M$ values obtained by random sampling the \citetalias{kroupa_imf} IMF while varying its parameters within observed uncertainties. While these runs have statistically indistinguishable IMFs (see Figure \ref{fig:unimportant_params_IMF}), their $L/M$ values vary significantly between runs. (Right) The \emph{final} $L/M$ values and $\Mmax$ values for all runs presented in \S\ref{sec:results_sensitivity}. 
The shaded region shows the 95\% confidence intervals for sampling the canonical \citetalias{kroupa_imf} IMF. $L/M$ and $\Mmax$ are highly correlated, so $L/M$ is probing the highest masses of the IMF, making it sensitive to sampling effects.}
\label{fig:LM_Mmax}
\vspace{-0.5cm}
\end {center}
\end{figure*}

We note that even though the IMF appears to vary mildly between the runs presented here, other summary statistics  of star formation, like the light-to-mass ratio $L/M$, can vary significantly. Figure \ref{fig:LM_Mmax} shows that although the set of runs shown in \ref{sec:sensitivity_alpha}-\ref{sec:sensitivity_mass} exhibit statistically indistinguishable IMFs, their final light-to-mass ratios vary by orders of magnitude. This is because $L/M$ is highly sensitive to the most massive stars due to the steep scaling of stellar luminosity with stellar mass. Overall $L/M\appropto \Mmax$, and their values fall within the 95\% confidence interval we obtain by sampling the \citetalias{kroupa_imf} IMF. We also find that in our simulations clouds with higher final total stellar mass have a higher $\Mmax$ and thus higher $L/M$. We do not have sufficient statistics to rule out the existence of a high-mass cut-off for the IMF that depends on initial conditions (see e.g., \citealt{weidner_2006_max_stellar_mass}). 

\section{Discussion}\label{sec:discussion}

A key goal of the STARFORGE project is to understand the roles different physical processes play in star formation by carrying out a suite of simulations with increasingly complex physics. A significant advantage of this suite, relative to comparing results from various groups in the literature, is that they use the same code base and initial conditions, allowing for a cleaner comparison. 

\subsection{Role of magnetic fields, thermodynamics and stellar feedback}

Our suite proceeded through the \myquote{physics ladder} of star formation, starting from isothermal MHD+gravity, then adding non-isothermal gas thermodynamics, protostellar jets, stellar radiation, winds and SNe. 

Similar to other recent work in the literature \citep[e.g., ][]{Haugbolle_Padoan_isot_IMF}, we found that in STARFORGE runs the magnetic fields impose a well-defined mass scale on the stellar mass spectrum \citepalias{Guszejnov_isoT_MHD}, preventing the runaway fragmentation found in non-magnetized, isothermal runs \citep{guszejnov_isothermal_collapse}. This mass scale, however, is an order of magnitude higher than observed for MW-like conditions. It is also sensitive to initial conditions (e.g., surface density) in a way that could not be reconciled with the apparent universality of the IMF in the MW \citep{imf_universality, guszejnov_imf_var}. 

Due to the highly efficient cooling of molecular gas, most star formation models assume the gas to be isothermal \citep{Girichidis_2020_sf_processes_review}. Detailed studies, however, showed that there is a significant scatter in the gas temperature, with a clear density dependence (see \citealt{Glover_Clark_n_T}). At high densities the isothermal assumption inevitably breaks down as the gas becomes opaque to its own cooling radiation, forming a hydrostatic Larson core \citep{larson_1969}. This transition to an adiabatic behavior can impose a mass scale on the stellar mass spectrum \citep{lowlyndenbell1976, rees1976}, but it was found to be significantly below the observed characteristic stellar mass scales. But recent works proposed that tidal screening around Larson cores can raise the relevant mass scales to be close to observed values \citep{Lee_Hennebelle_2018_EOS,Colman_Teyssier_2019_tidal_screening}. Note that these works all pertain to non-magnetized clouds without feedback, so the only unique mass scale is imposed by the isothermal-adiabatic transition. In \citetalias{guszejnov_starforge_jets} we investigated these effects for magnetized clouds and found that transitioning to non-isothermal thermodynamics had little effect on the stellar mass spectrum, which was still predominantly set by magnetic effects, leading to stellar masses significantly above those observed. In this work we carried out RMHD simulations where heating and cooling radiation is explicitly followed, unlike in \citetalias{guszejnov_starforge_jets}. This allows the gas to self-shield, leading to temperatures below the 10 K that is the fiducial temperature in isothermal approximations and was used as a floor temperature in \citetalias{guszejnov_starforge_jets}. This leads to a reduction of stellar masses, but the overall mass scales (with only these physics) remain significantly above the observed values (see Figure \ref{fig:imf_compare}). 

Previous work in the literature has shown that protostellar jets significantly affect star formation. Jets directly reduce stellar masses and slow down star formation in the cloud  \citep{Cunningham_2011_outflow_sim, hansen_lowmass_sf_feedback,Federrath_2014_jets} and can potentially drive turbulence on small scales \citep[e.g.,][]{Nakamura_2007_outflow_turbulence_driving,Wang_2010_outflow_regulated_SF,Offner_Arce_2014,Offner_Chaban_2017_jets_sfe, murray_2018_jets}. In \citetalias{guszejnov_starforge_jets} we showed that protostellar jets play a vital role in setting stellar masses, reducing them by an order of magnitude. This reduction is significantly larger than what one would expect if the jets simply removed of order $1/3$ of the accreted material, as jets not only remove accreted material but also disrupt the accretion flows around protostars, allowing the nearby ISM to fragment and form new stars. Overall, in \citetalias{guszejnov_starforge_jets} we found that jets bring the stellar mass spectrum in line with observations in the MW, with the exception of the most massive stars. In this work we reran the same simulations with explicitly evolved radiative heating and cooling (RHD) and found the results to be largely the same. In both works protostellar jets significantly affect the virial state of the cloud and can suppress the formation of new stars. However, they are unable to expel the remaining gas and to prevent massive stars from accreting it, leading to runaway accretion in high-mass stars.

Radiative feedback from stars has long been theorized to play an important role in setting stellar mass scales (e.g., \citealt{krumholz_stellar_mass_origin,bate12a,Myers_2013_ORION_radiation_IMF, guszejnov_feedback_necessity,li_2018_sf_mhd_jets}, see \citealt{Hennebelle_2020_IMF_stellar_feedback} for a counterexample) and regulating star formation \citep[e.g., ][]{Offner_2009_radiative_sim, krumholz_2012_orion_sims, Cunningham_2018_feedback}. Many of these previous works have produced mass spectra similar to the observed IMF, generally by including most of the relevant feedback processes (jets, radiation) in magnetized clouds \citep[e.g, ][]{Cunningham_2018_feedback}. This work expands upon these results in several ways. First, the clouds in our simulations are an order of magnitude more massive than those in previous works with similar physics \citep[e.g., ][]{Cunningham_2018_feedback}. Second, the \myquote{physics ladder} suite allows us to disentangle the effects of the individual feedback processes. We see that radiative feedback plays a key role in disrupting the cloud and quenching star formation. In all our simulations the formation of the first main sequence O star marks a turning point in the global evolution of the cloud and the beginning of the disruption process. Note that most of the aforementioned previous works in the literature simulated much smaller clouds ($200-1000\,\msun$), so no massive stars formed in them. Stellar winds further enhance feedback from massive stars, but do not significantly alter it as they are only a significant channel of momentum feedback for massive O stars. Radiative feedback and winds both counteract the runaway accretion onto massive stars found in jet runs, both by shutting off accretion and by generally expelling gas from the cloud. Therefore, jets are responsible for setting the peak of the IMF but radiative and wind feedback are responsible for preventing the high-mass end of the IMF from flattening. Note, however that even with stellar feedback the resulting high-mass IMF slopes are consistent with the -2 value expected from scale-free fragmentation.

Finally, massive stars end their lives as supernovae (SNe). These explosions are generally agreed to be critical for regulating star formation on galactic scales, and in particular to dominate the overall momentum input in the ISM by stellar feedback \citep{Somerville_Dave_2015_galaxy_formation_review, naab_ostriker_galform_review, vogelsberger_galform_review}, although non-linear interactions between different processes are also important \citep[e.g.,][]{hopkins2014_fire,Hopkins_2018_sne_feedback}. However, they occur fairly late in the star formation process; simulations of cluster formation have found they have negligible effects upon SFE and bound cluster masses compared to early feedback, even in massive GMCs that survive long enough to host SNe before disruption \citep{grudic_2020_cluster_formation}. In this work we similarly find that SNe occur after the cloud has been completely disrupted by earlier feedback processes. In almost all of our simulations the cloud is disrupted within 0.5-1.0 freefall times after the first O star forms, the only exception being the $\mu=0.4$ highly magnetized run, which does not completely disrupt until the first SNe go off. Note, that for massive clouds ($10^5-10^6\,\msun$), we could plausibly expect SNe to trigger before the cloud is disrupted, even for less magnetized clouds. Even if SNe turn out to have no direct role in regulating SF in massive clouds, they are likely to have a major indirect role as they are thought to be one of the main drivers of galactic turbulence and thus set the properties of GMCs \citep[e.g.,][]{ostriker_shetty_sf_turbulence_sne,Hopkins_2011_SFR_self_regulate,hopkins_2012_galaxy_structure,Faucher_Giguere_2013_SF_feedback_galactic_disks, Walch_2015_SILCC_ISM_SN,Martizzi_2016_SN_feedback_ISM, Padoan_2017_SN_driving_SFE,Seifried_2018_GMC_SN_driving, guszejnov_GMC_cosmic_evol,Gurvich_2020_ISM_pressure_balance__gal_disks}.

\subsection{Environmental variations}

In \S\ref{sec:results_sensitivity} we analysed the star formation history and stellar mass spectrum in simulations that include all levels of feedback physics (\textbf{C\_M\_J\_R\_W}) while varying various initial cloud parameters (see \citetalias{Guszejnov_isoT_MHD} and \citetalias{guszejnov_starforge_jets} for similar studies for the lower rungs of the \myquote{physics ladder}). 

We find that the final star formation efficiencies of the clouds significantly depend on initial cloud parameters, specifically the initial surface density, level of turbulence, magnetization and metallicity (i.e., gas temperature; see Eq. \ref{eq:fit_SFE}), varying between 1\% and 12\% in the simulated clouds. These trends are in agreement with expectations from simple rule-of-thumb considerations, such as higher surface density making it harder for feedback to unbind the cloud, thus leading to a higher $\SFE$, just as increased initial turbulence or magnetic support makes it easier to unbind the cloud, lowering the final $\SFE$. Lowering the metallicity of the initial gas or lowering the ISRF lowers the final SFE, although the exact mechanism is unclear.

Regarding the sink mass spectrum (IMF), the initial cloud surface density and virial parameter have little effect on the final median stellar mass scale (see Eq. \ref{eq:fit_mmedian}). This appears to contradict \citetalias{Guszejnov_isoT_MHD} and \citetalias{guszejnov_starforge_jets} where we found both parameters to significantly affect the stellar mass spectrum. The apparent contradiction is resolved by taking into account that those works lacked the relevant feedback physics to quench star formation and disrupt the cloud, thus all comparisons were done at fixed $\SFE$. For fixed $\SFE$ the runs presented in this work show similar variations. However, when comparing the \emph{final}, post-disruption IMFs, we find that the dependence of the final $\SFE$ on cloud properties effectively cancels these variations. Insensitivity to both the $\alphaturb$ virial parameter and $\Sigma$ surface density is required if we are to reproduce the near-universal IMF of the MW \citep{imf_universality}, as observed molecular clouds in similar regions exhibit an order of magnitude scatter in $\alphaturb$ and a factor of 5 in $\Sigma$ \citep{heyer_2009_larson,kauffmann_pillai_2013, heyer_dame_2015}. Following Table \ref{tab:scalings}, we predict mild variations less than few tens of percents in $\Mmedian$, well within the observational uncertainties of the MW IMF. Decreasing metallicity or increasing the local ISRF raises the gas temperature, which in turn increases the relevant mass scales of star formation (e.g., Jeans mass, sonic mass). This leads to an increase of the median stellar mass, however the shift is very small, roughly consistent with $\Mmedian\appropto Z^{-1/10} e_\mathrm{ISRF}^{1/10}$. This is consistent with the weak trend predicted by \citet{Sharda_Krumholz_2021_IMF_bottom_heavy_Z} for the characteristic stellar mass for $Z>0.01\,Z_\odot$. Previous simulations that included only a subset of the physics presented here (e.g., no MHD, jets or winds) found similarly no significant IMF variations with metallicity \citep{Bate_2019_Z_multiplicity}. Overall $\Mmedian$ varies very weakly with initial gas parameters, consistent with the observed limited variations in the stellar IMF. The high-mass slope of the IMF is more sensitive to initial conditions, steepening with increasing initial magnetization and becoming more shallow for lower metallicity values. 

Although our simulations only tested the effects of mild variations in initial parameters, we can extrapolate them to the more extreme star forming regions, such as the Central Molecular Zone of the MW, starburst galaxies, or high-redshift galaxies. These regions have surface densities a factor $100-1000$ higher than in the MW \citep{Solomon_1997_ULIRG_ISM, Swinbank_2011_dense_galaxy_ISM} and an ISRF that is a factor 100x higher. While we plan to simulate star formation in such environments in the future, for now we can make a rough estimate of the IMF with Table \ref{tab:scalings} and find $\Mmedian$ to be within a factor of 2 of the MW value. 

\subsection{Caveats}\label{sec:caveats}

While the simulation presented here are the current state-of-the-art for simulating star forming clouds, like other simulations in the literature STARFORGE employs a large number of significant approximations and assumptions to make the simulations computationally tractable (see \citetalias{grudic_starforge_methods} for detailed discussions). In particular, the runs used here have a $\sim 30\,\mathrm{AU}$ Jeans-resolution, i.e. fragmentation on scales smaller than this are not resolved. This has a dramatic effect on the formation of protostellar disks and their fragmentation, causing the simulation to potentially miss closely formed binaries and overestimate stellar masses. However, we do not expect it to qualitatively affect the IMF above $\Mcompleteness$ (see Appendix \ref{app:completeness}), except for potentially steepening the high-mass slope as massive stars are broken up through disk fragmentation. 

Recent observations of dwarf galaxies \citep{HunterDA_2021_SFR_velocity_dispersion_corr, Elmegreen_2022_SFR_velocity_dispersion_corr, HunterLC_2022_SFR_velocity_dispersion_corr} showed that on galactic scales ($\sim 400\pc$) the ISM velocity dispersion correlates best with the local star formation rate with a 100 Myr delay. Meanwhile in our simulations clouds are destroyed in about 2 freefall times (corresponding to roughly 10 Myr) and the kinetic energy of the gas increases by an order of magnitude (see Figure \ref{fig:disruption_energy_evol}). Due to the relatively small size of the simulated volume and the simplified modeling of the surrounding ISM, the distribution and dissipation of kinetic energy in the ISM after cloud disruption is not captured accurately, and will be revisited in future work. 

\section{Conclusions}\label{sec:conclusions}

In this work we presented simulations from the STARFORGE project, which are high resolution radiation-MHD simulations following the evolution of star forming molecular clouds. The runs include progressively more complex physics, starting from isothermal MHD, then enabling explicitly solved heating and cooling radiation and adding stellar feedback in the form of protostellar jets, radiation, stellar winds and supernovae. Building on our past work we investigate each rung of this \myquote{physics ladder} of star formation to identify the role each process plays in star formation.

In previous works we showed that isothermal MHD leads to a well-defined stellar mass spectrum \citepalias{Guszejnov_isoT_MHD}, and that the addition of protostellar jets is necessary to bring these scales in line with observations \citepalias{guszejnov_starforge_jets}. The runs presented in this paper reinforce those conclusions: stellar mass scales are set by MHD turbulence that both creates the self-gravitating structures and prevents their runaway fragmentation (see non-magnetized case in \citealt{guszejnov_isothermal_collapse}). Protostellar jets dramatically reduce stellar mass scales by both directly removing accreted material and by disrupting the accretion flow around stars, however they cannot prevent the most massive stars from undergoing runaway accretion. In these runs radiation was explicitly evolved, allowing gas to cool below the isothermal temperature limit in dense regions. This leads to a significant reduction in stellar masses, which was not captured in \citetalias{guszejnov_starforge_jets}. The addition of stellar radiation, winds and supernovae have little direct effect on the stellar mass spectrum, apart from preventing the runaway accretion of massive stars. They, however, play a dominant role in regulating star formation. In the presented runs stellar radiation and protostellar jets are the dominant forms of feedback that quench star formation and disrupt the cloud, with the formation of the first main sequence O star marking the turning point in the cloud's evolution. While supernovae do go off in these simulations, these exclusively happen at the end of the runs when the cloud has already been disrupted by radiative feedback. It should be noted that our simulations followed $\le 2\times 10^4\,\msun$ clouds with lifetimes of $\sim 7\,\mathrm{Myr}$, so it is possible that SNe play a significant role in more massive clouds whose lifetimes are longer, which we plan to explore in future work.

In addition to the \myquote{physics ladder} suite, we present a suite of full physics simulations with varied initial parameters to determine how the star formation history and stellar mass spectrum (IMF) depend on initial conditions. The characteristic stellar masses are insensitive to the initial cloud mass, surface density and level of turbulence. Note that this only applies to the final, post-disruption mass spectrum; comparisons at fixed times or star formation efficiencies show significant differences. Of the parameters probed in this study, the IMF peak is only affected by the cloud metallicity and the strength of the interstellar radiation field (ISRF). Since both significantly alter the thermodynamics of the cloud, we conjecture that their effects can be attributed to a change in the mean temperature of star-forming gas. Meanwhile, the high-mass slope of the IMF becomes steeper with decreasing metallicity or increasing  ISRF and magnetization. The scaling relations derived from our parameter study predict IMF variations that are within the observational uncertainties of the near-universal IMF observed in the MW.
  
 

\section*{Acknowledgements}
DG is supported by the Harlan J. Smith McDonald Observatory Postdoctoral Fellowship and the Cottrell Fellowships Award (\#27982) from the Research Corporation for Science Advancement. 
Support for MYG was provided by NASA through the NASA Hubble Fellowship grant \#HST-HF2-51479 awarded  by  the  Space  Telescope  Science  Institute,  which  is  operated  by  the   Association  of  Universities  for  Research  in  Astronomy,  Inc.,  for  NASA,  under  contract NAS5-26555.
SSRO was supported by NSF CAREER Award AST-1748571, NASA grant 80NSSC20K0507, NSF grant 2107942 and by a Cottrell Scholar Award (\#24400) from the Research Corporation for Science Advancement. 
CAFG was supported by NSF through grants AST-1715216, AST-2108230,  and CAREER award AST-1652522; by NASA through grants 17-ATP17-006 7 and 21-ATP21-0036; by STScI through grants HST-AR-16124.001-A and HST-GO-16730.016-A; by CXO through grant TM2-23005X; and by the Research Corporation for Science Advancement through a Cottrell Scholar Award. 
Support for PFH was provided by NSF Collaborative Research Grants 1715847 \&\ 1911233, NSF CAREER grant 1455342, and NASA grants 80NSSC18K0562 \&\ JPL 1589742.
ALR  acknowledges support from Harvard University through the ITC Post-doctoral Fellowship.
This work used computational resources provided by XSEDE allocations AST-190018 and AST-140023, the Frontera allocation AST-20019, and additional resources provided by the University of Texas at Austin and the Texas Advanced Computing Center (TACC; http://www.tacc.utexas.edu).

\section{Data availability}
The data supporting the plots within this article are available on reasonable request to the corresponding authors. Additional figures can be found at our GitHub repository \url{https://github.com/guszejnovdavid/STARFORGE_IMF_paper_extra_plots}. A public version of the {\small GIZMO} code is available at \url{http://www.tapir.caltech.edu/~phopkins/Site/GIZMO.html}.

 


 \bibliographystyle{mnras}
 \bibliography{bibliography} 



\appendix

\section{Resolution effects on the sink mass spectrum}\label{app:completeness}

The fiducial resolution level in our simulations is $\dderiv m=10^{-3}\,\msun$ (equivalent to $\dderiv x_\mathrm{Jeans}\sim 20\,\AU$, see \citetalias{grudic_starforge_methods}), a choice based on our previous work in \citetalias{Guszejnov_isoT_MHD} and \citetalias{guszejnov_starforge_jets}, where this value was sufficient for the sink mass spectrum to be complete down to $\Mcompleteness=0.1\,\msun$. With the transition to the new RHD based thermodynamics module and the inclusion of stellar radiation and wins, it is worth reexamining this choice. We do so by running a \myquote{full physics} (\textbf{C\_M\_J\_R\_W}) simulation at different resolution levels. Since RHD simulations with $M_0/\dderiv m\gg 2\times 10^{7}$ are prohibitively expensive, we choose to do this resolution study on a smaller \textbf{M2e3} cloud (see Table \ref{tab:IC_phys}) within a resolution range of $\dderiv m \in [10^{-2},10^{-4}]$.

\begin{figure*}
\begin {center}
\includegraphics[width=0.33\linewidth]{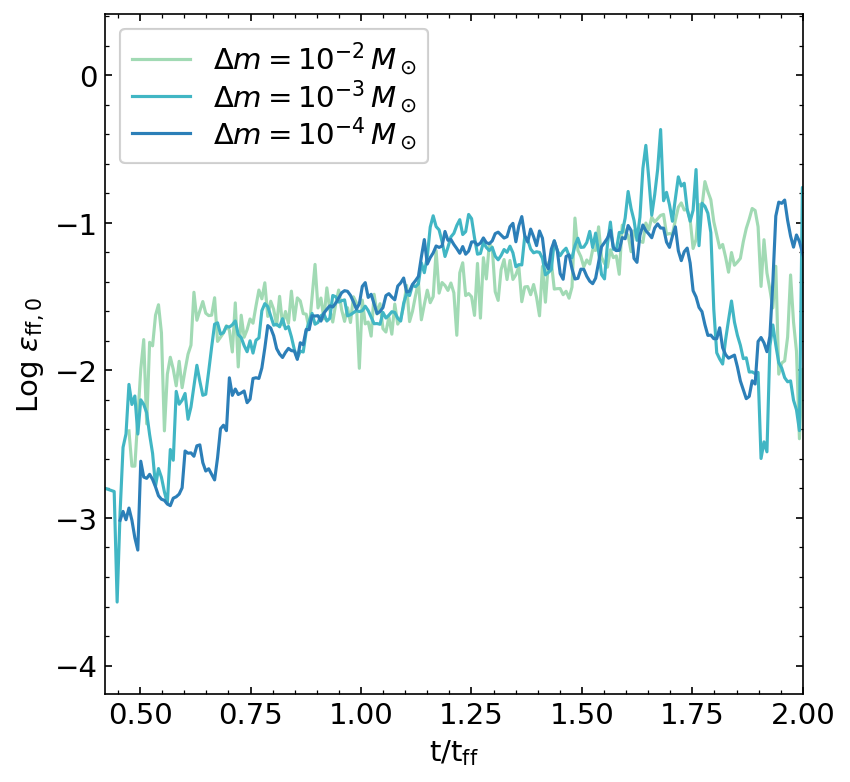}
\includegraphics[width=0.33\linewidth]{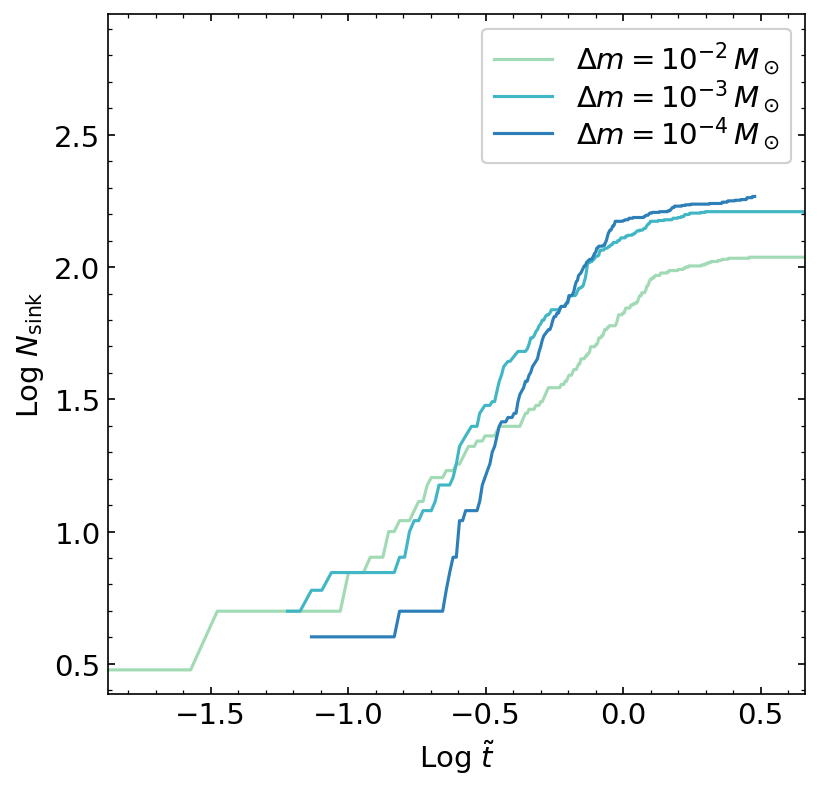}
\includegraphics[width=0.33\linewidth]{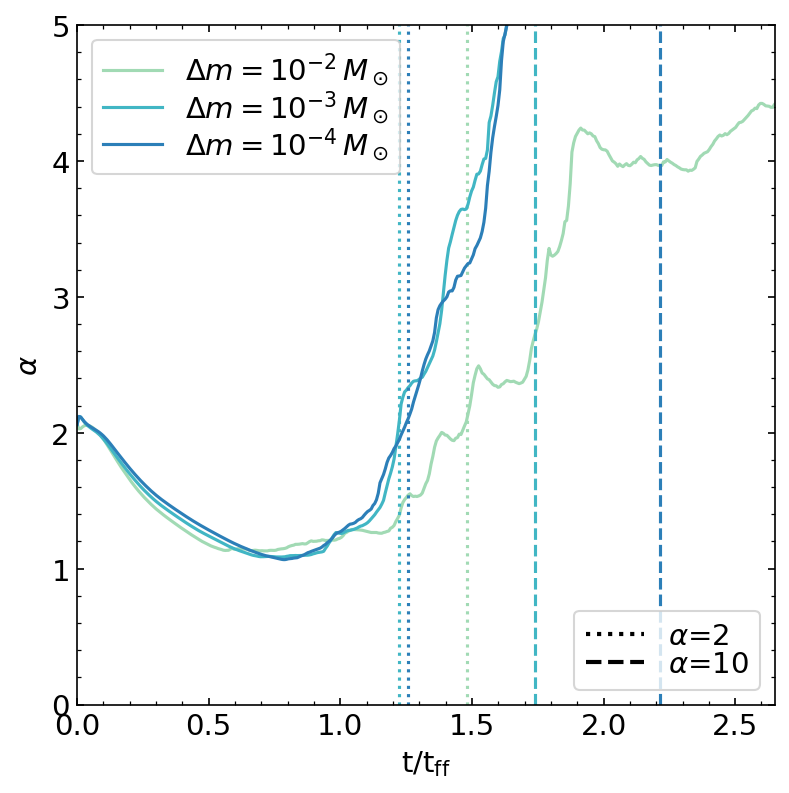}
\vspace{-0.4cm}
\caption{The evolution of the star formation rate per freefall time ($\epsff$, left), number of sink particles ($\Nsink$, middle) and the virial parameter ($\alpha$, right) as a function of time for an \textbf{M2e3} cloud with all physics included (\textbf{C\_M\_J\_R\_W}, see Table \ref{tab:IC_phys}).}
\label{fig:resolution_test_SF}
\vspace{-0.5cm}
\end {center}
\end{figure*}

Figure \ref{fig:resolution_test_SF} shows that the star formation history of the cloud is insensitive sensitive to numerical resolution in the examined range. The final sink particle number and the evolution of the cloud virial state are virtually identical between our fiducial resolution and the 10 times higher value.

\begin{figure*}
\begin {center}
\includegraphics[width=0.45\linewidth]{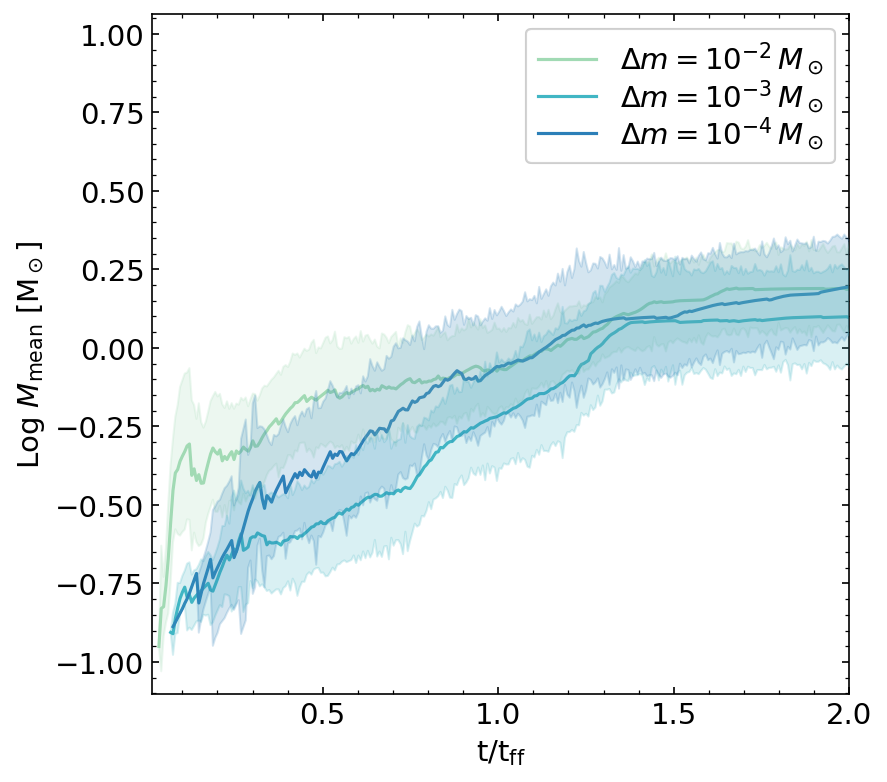}
\includegraphics[width=0.45\linewidth]{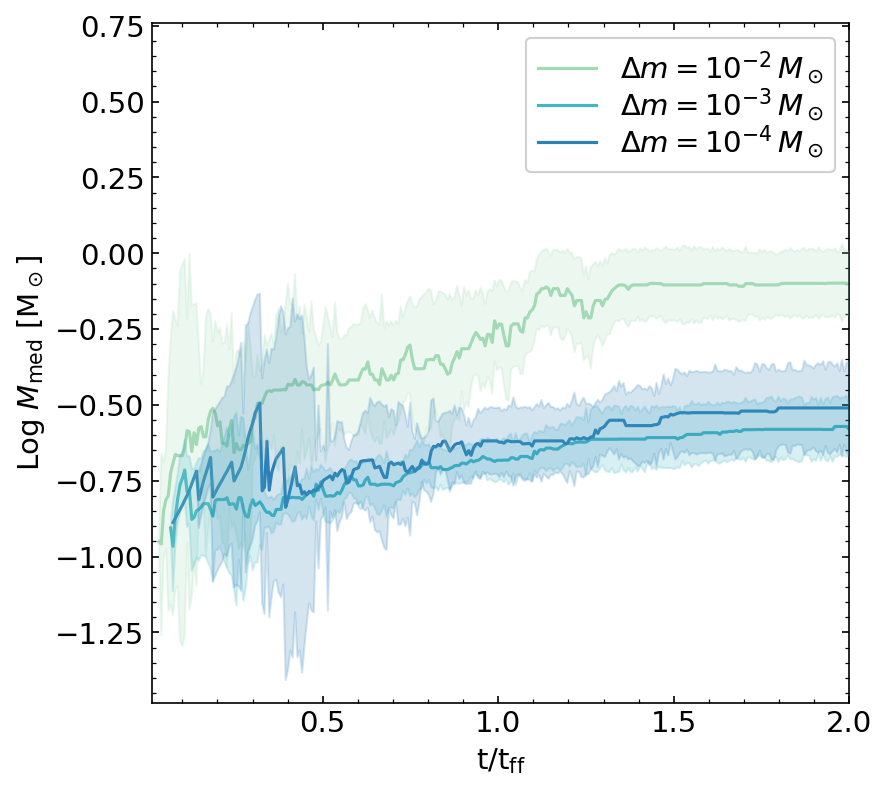}\\
\includegraphics[width=0.45\linewidth]{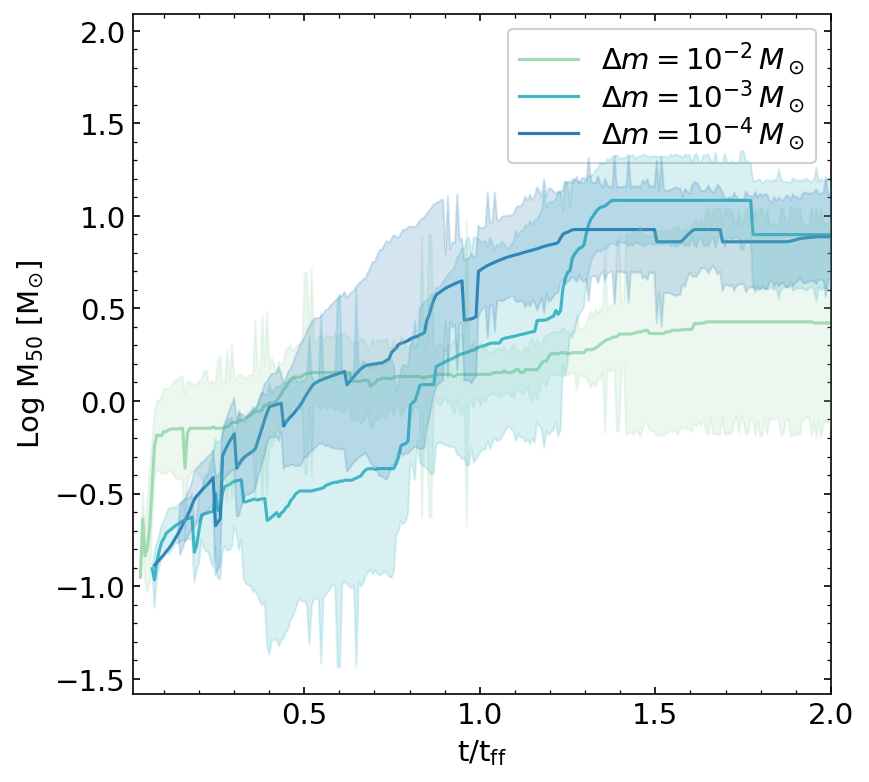}
\includegraphics[width=0.45\linewidth]{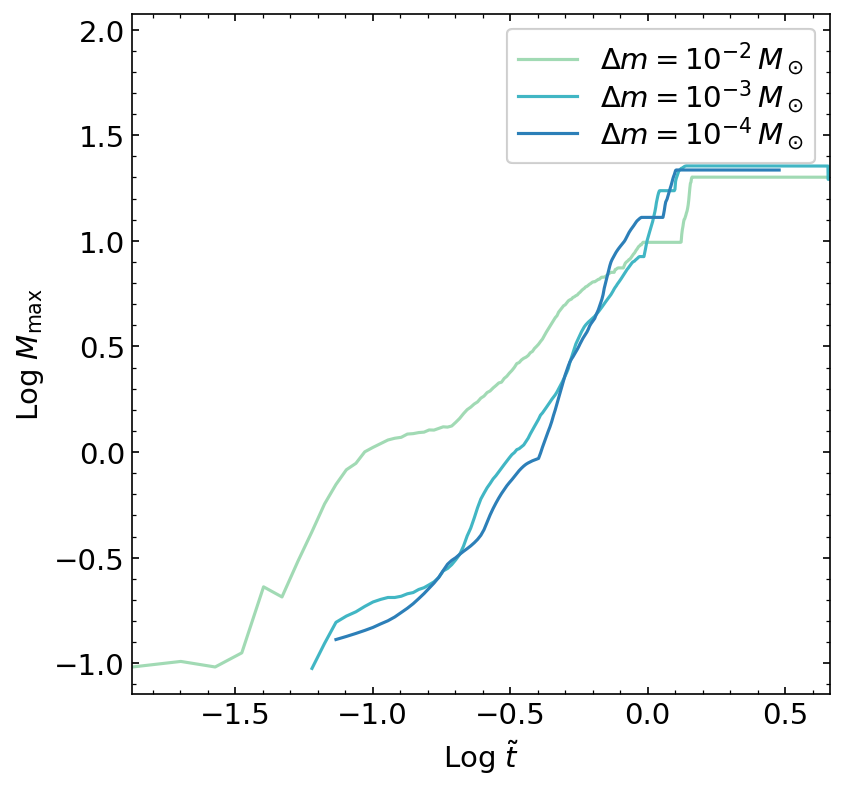}
\vspace{-0.4cm}
\caption{The evolution of the number-weighted mean ($\Mmean$, top left), number-weighted median ($\Mmedian$, top right), mass-weighted median ($\Mmassmedian$, bottom left) and maximum ($\Mmax$, bottom right) sink mass as a function of time for an \textbf{M2e3} cloud with all physics included (\textbf{C\_M\_J\_R\_W}, see Table \ref{tab:IC_phys}). Shaded regions show the 95\% confidence intervals. Note that all mass scales are calculate for sinks above our chosen limit of $\Mcompleteness=0.1\,\msun$.}
\label{fig:resolution_test_masses}
\vspace{-0.5cm}
\end {center}
\end{figure*}

Figure \ref{fig:resolution_test_masses} shows that the mean, median and maximum sink masses are essentially identical between the fiducial $\dderiv m=10^{-3}\,\msun$ and the higher resolution run. Note that these metrics are calculated above the same $\Mcompleteness=0.1\,\msun$ as in the rest of the paper, regardless of the mass resolution of the simulation.

\begin{figure}
\begin {center}
\includegraphics[width=0.99\linewidth]{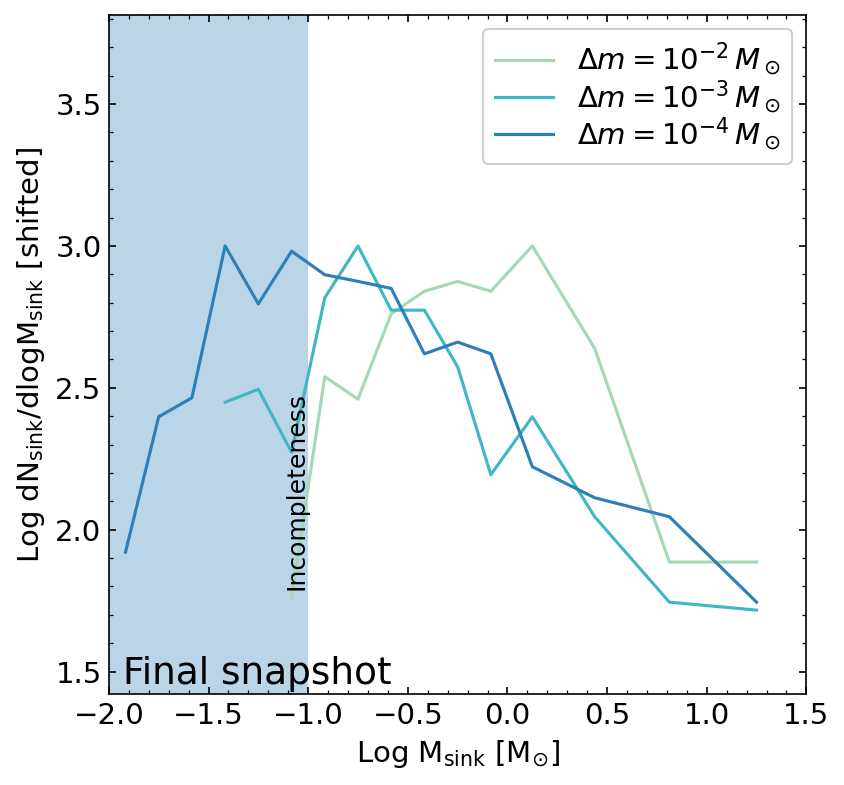}
\vspace{-0.4cm}
\caption{The sink mass spectrum (IMF) for an \textbf{M2e3} cloud with all physics included (\textbf{C\_M\_J\_R\_W}, see Table \ref{tab:IC_phys}) at different $\Delta m$ mass resolutions.}
\label{fig:resolution_test_imf}
\vspace{-0.5cm}
\end {center}
\end{figure}

Figure \ref{fig:resolution_test_imf} shows the mass distribution of stars at the end of the simulations. As expected, the spectrum extends to lower masses with higher resolution, however the part above $\sim 0.1\,\msun$ to be identical between the fiducial and the high resolution run. This is similar to the results we obtained in our previous works \citepalias{guszejnov_starforge_jets}, leading to a conservative, rule-of-thumb estimate of our completeness limit as $\Mcompleteness\sim 100\dderiv m$. For our fiducial resolution this leads to $\Mcompleteness=0.1\msun$, which we adopt as our completeness limit for this work.

\bsp	
\label{lastpage}
\end{document}